\documentclass[aip,pof,reprint]{revtex4-1}

\usepackage{graphicx}
\usepackage{amsmath, amssymb, amsfonts}
\usepackage{subcaption}
\PassOptionsToPackage{hyphens}{url}\usepackage{hyperref}
\usepackage{color}
\usepackage{siunitx}
\graphicspath{{../Figures/}}

\newcommand\Rey{\mbox{\textit{Re}}}
\newcommand\Bin{\mbox{\textit{Bi}}}
\newcommand\HB{\mbox{\textit{HB}}}
\newcommand\Pap{\mbox{\textit{Pa}}}

\newcommand{\sfrac}[2]{\mathchoice
  {\kern0em\raise.5ex\hbox{\the\scriptfont0 #1}\kern-.15em/
   \kern-.15em\lower.25ex\hbox{\the\scriptfont0 #2}}
  {\kern0em\raise.5ex\hbox{\the\scriptfont0 #1}\kern-.15em/
   \kern-.15em\lower.25ex\hbox{\the\scriptfont0 #2}}
  {\kern0em\raise.5ex\hbox{\the\scriptscriptfont0 #1}\kern-.2em/
   \kern-.15em\lower.25ex\hbox{\the\scriptscriptfont0 #2}}
  {#1\!/#2}}
\newcommand{\myhalf}{\sfrac{1}{2}}
\newcommand{\nph}{{m + \myhalf}}
\newcommand{\nmh}{{m - \myhalf}}
\newcommand{\dt}{\Delta t}
\newcommand{\bu}{\boldsymbol{u}}
\newcommand{\G}{\boldsymbol{G}}
\newcommand{\D}{\boldsymbol{D}}

\DeclareMathAlphabet{\mathsfbi}{OT1}{\sfdefault}{bx}{sl}
\DeclareMathVersion{sfletters}
\SetSymbolFont{letters}{sfletters}{OML}{ntxsfmi}{b}{it}

\makeatletter
\newcommand{\mathbfsbilow}[1]{%
  \text{\mathversion{sfletters}$\m@th#1$}%
}

\DeclareRobustCommand{\tensor}[1]{%
  \begingroup
  \ifcat\noexpand #1\relax
    \edef\greek@test{\detokenize{#1}}%
    \edef\greek@test{\expandafter\@cdr\greek@test\@nil}%
    \edef\greek@test{\expandafter\@car\greek@test\@nil}%
    \edef\x{\the\lccode\expandafter`\greek@test}%
    \edef\y{\number\expandafter`\greek@test}%
    \ifnum\x=\y\relax
      \mathbfsbilow{#1}%
    \else
      \mathsfbi{#1}%
    \fi
  \else
    \mathsfbi{#1}%
  \fi
  \endgroup
}
\makeatother

\draft 

\begin{document}


\title{Highly parallelisable simulations of time-dependent viscoplastic fluid flow simulations with
structured adaptive mesh refinement}



\author{Knut Sverdrup}
\email{ksk38@cam.ac.uk}

\author{Nikolaos Nikiforakis}
\affiliation{Cavendish Laboratory, University of Cambridge, CB3 0HE, UK}

\author{Ann Almgren}
\affiliation{Lawrence Berkeley National Laboratory, CA 94720, USA}

\date{\today}

\begin{abstract}
We present the extension of an efficient and highly parallelisable framework for incompressible
fluid flow simulations to viscoplastic fluids.  The system is governed by incompressible
conservation of mass, the Cauchy momentum equation and a generalised Newtonian constitutive law. In
order to simulate a wide range of viscoplastic fluids, we employ the Herschel-Bulkley model for
yield-stress fluids with nonlinear stress-strain dependency above the yield limit. We utilise
Papanastasiou regularisation in our algorithm to deal with the singularity in apparent viscosity.
The resulting system of partial differential equations is solved using the IAMR code
(Incompressible Adaptive Mesh Refinement), which uses second-order Godunov methodology for the
advective terms and semi-implicit diffusion in the context of an approximate projection method to
solve on adaptively refined meshes. By augmenting the IAMR code with the ability to simulate
regularised Herschel-Bulkley fluids, we obtain efficient numerical software for time-dependent
viscoplastic flow in three dimensions, which can be used to investigate systems not considered
previously due to computational expense. We validate results from simulations using this new
capability against previously published data for Bingham plastics and power-law fluids in the
two-dimensional lid-driven cavity. In doing so, we expand the range of Bingham and Reynolds numbers
which have been considered in the benchmark tests.  Moreover, extensions to time-dependent flow of
Herschel-Bulkley fluids and three spatial dimensions offer new insights into the flow of
viscoplastic fluids in this test case, and we provide missing benchmark results for these
extensions.
\end{abstract}


\maketitle 




\section{Introduction}
\label{sec:introduction}

Viscoplastic fluids are non-Newtonian fluids which are characterised by a minimum induced stress
necessary for flow to occur. For this reason they are also commonly referred to as yield-stress
fluids. When the imposed stress does not exceed the threshold value, the material is modelled as a
rigid solid. In regions where the yield stress is exceeded, however, the material flows like a
fluid. The ability of the material to support a stress under certain circumstances gives rise to
phenomena such as non-flat surfaces at rest under gravity and the coexistence of yielded (flowing)
and unyielded (rigid) regions within the fluid.  The former can be demonstrated by distorting the
surface of mayonnaise in a jar: gravity alone is not strong enough to surpass the yield stress, and
the surface remains in its distorted state. In addition to being fundamentally interesting from the
perspectives of rheology, fluid mechanics and mathematical modelling, yield stress fluids occur
naturally and are paramount to the success of animals such as mudskippers
\cite{pegler2013locomotion} and snails \cite{denny1981quantitative}.  Their importance in
industries ranging from medicine \cite{apostolidis2014modeling, apostolidis2015modeling,
apostolidis2016effect, apostolidis2016non} to oil and gas exploration \cite{bittleston1991mud,
taghavi2012incomplete, frigaard2017bingham} has led to extensive research contributions in the
field.

Just as for Newtonian fluids, the pursuit of knowledge about viscoplastic fluids has relied heavily
on computational methods in the last fifty years. Compared to the Newtonian case, however, the
numerical simulations are much more demanding in terms of processing time. This is largely due to
the existence of a yield stress, since this results in a singularity at zero strain rate for the
apparent viscosity of the fluid. Traditional methods for computing the solution are rendered useless
for this case, since infinite viscosities cannot be represented in the unyielded regions. There are
now two main branches of algorithms designed to deal with this problem. The first utilises
mathematical regularisation to approximate the viscosity function in the low-strain limit. In doing
so, the rigid body approximation is effectively replaced by a fluid with very high viscosity in the
unyielded regions. A regularisation parameter controls how close this approximation is to the
idealised viscoplastic case. On the other hand, the problem can be reformulated in the framework of
non-smooth optimisation theory, and solved using augmented Lagrangians. Such methods can solve the
unregularised problem without introducing any approximations, but generally require more
computational resources to find the solution \cite{saramito2017progress}. 

The regularisation approach was first explored by Bercovier and Engelman in 1980
\cite{bercovier1980finite}, who utilised a simple yet efficient work-around by adding a small
constant to the computed strain-rate, so that even in the zero-strain limit the viscosity would
remain finite. An alternative method proposed by Tanner and Milthorpe a few years later was the
bi-viscosity model \cite{tanner1983numerical}, in which the viscoplastic fluid is characterised by
a separate, large viscosity when the strain-rate is below a given threshold (see e.g.~Abhijit and
Sayantan \cite{guha2016analysis}). An exponential regularisation factor was then introduced by
Papanastasiou in 1987 \cite{papanastasiou1987flows}, and this method of regularisation is still
widely used in modern codes which deal with viscoplasticity through regularisation. Important
investigations based on Papanastasiou regularisation include those of Mitsoulis et
al.~\cite{mitsoulis2001flow, zisis2002viscoplastic, mitsoulis2004creeping, mitsoulis2017numerical}
and Syrakos et al.~\cite{syrakos2013solution, syrakos2014performance, syrakos2016cessation}.

Augmented Lagrangian methods were first applied to the viscoplastic flow problem in 1983 by Fortin
and Glowinski \cite{fortin1983chapter}, but the variational formulation on which it relies was
derived by Duvaut and Lions in 1976, who studied existence, uniqueness and regularity of solutions
to the problem \cite{duvaut1976inequalities}. The augmented Lagrangian algorithm itself is due to
Hestenes (1969) \cite{hestenes1969multiplier}.  Although the simulation of viscoplastic flow using
this optimisation technique constitutes an important milestone in the history of its numerical
treatments, the regularisation approach was much more popular due to the large discrepancy in
computational resource requirements. Advances in convex optimisation over the next few decades led
to the work of Saramito and Roquet \cite{saramito2001adaptive}, where significant improvements were
achieved in terms of convergence rates, and hence, computational efficiency. Other contributions
\cite{vola2003laminar} further confirmed the potential, and in recent years state-of-the-art
algorithms are actively being developed, notably by Treskatis et
al.~\cite{treskatis2016accelerated}, Saramito \cite{saramito2016damped}, Bleyer
\cite{bleyer2018advances} and Dimakopoulos et al.~\cite{dimakopoulos2018pal}. Due to the ability to
solve the unregularised viscoplasticity problem, these methods have become increasingly popular
since algorithmic advances have led to significant speed-up of their runtime. It is important to
note, however, that these methods are still more costly than regularised approaches, so although
they allow computation of exact locations of yield surfaces, they are unnecessarily expensive when
this is less important, and a general understanding of the flow field is desirable. An example is
within cement displacement complexities for engineering purposes \cite{enayatpour2017advanced},
where the desirable insight is pressure distributions and the cement displacement efficiency
\cite{tardy2017new}.

Numerical treatment of viscoplastic flow problems has significantly improved over the last decades.
Researchers interested in analysing such flows are, however, still limited by the computational
cost associated with their solution. Notably, most research published in the field considers only
two spatial dimensions and only steady-state solutions. Although open-source libraries such as
FEniCS \cite{alnaes2015fenics} and OpenFOAM \cite{jasak2007openfoam} can be used to simulate
regularised viscoplastic fluids in three dimensions, our contribution aims to provide a massively
parallelisable tool for contemporary supercomputer architectures, with capabilities for structured
AMR (Adaptive Mesh Refinement). In order to achieve this, we begin with the IAMR
\cite{almgren1998conservative} code, that uses a second-order accurate, approximate projection
method to solve the incompressible Navier-Stokes equations.  The code is built on the AMReX
software framework for developing parallel block-structured AMR applications.  We have implemented
the Papanastasiou regularised Herschel-Bulkley model for the apparent viscosity function in IAMR so
that it can be used to simulate generalised Newtonian fluids. As such, we are able to take
advantage of modern supercomputer architectures in our simulation of viscoplastic fluids. For the
first time, results are presented for the transient, three-dimensional lid-driven cavity problem
for viscoplastic fluids.  Notably, the evolution of the three-dimensional yield surface from rest
to steady-state is tracked, before the lid velocity is set to zero, allowing cessation of the
viscoplastic fluid.

For further reading on developments in viscoplastic fluids, we refer the reader to the review
papers by Barnes \cite{barnes1999yield} and Balmforth et al.~\cite{balmforth2014yielding}.
Additionally, two excellent review papers on advances in numerical simulations of these fluids
appeared in a recent special edition of \emph{Rheologica Acta} \cite{mitsoulis2017numerical,
saramito2017progress}. In section \ref{sec:mathematical}, we will introduce the governing partial
differential equations and discuss relevant fluid rheologies. Section \ref{sec:numerical} is
devoted to the numerical algorithm employed to simulate the fluid flow, including the
regularisation strategy. Thorough validation is performed in section \ref{sec:validation},
before we evaluate the code for Herschel-Bulkley fluids and three dimensions in section
\ref{sec:evaluation}. Section \ref{sec:conclusions} concludes the article.


\section{Mathematical formulation}
\label{sec:mathematical}

Our domain $\Omega \subset \mathbb{R}^d$ is either in two or three dimensions, i.e.~$d \in
\{2,3\}$. For rank-2 tensors, we prescribe the scaled Frobenius norm
\begin{equation}
    |\tensor{A}| = \sqrt{\frac{1}{2}{\rm tr}(\tensor{A A}^\top)},
\end{equation}
as is customary in viscoplastic fluid mechanics.  We take the gradient of a vector $\boldsymbol{u}$
as the tensor with components $(\nabla \boldsymbol{u})_{ij} = \partial u_j / \partial x_i$, and the
symmetric part of this gradient is given by $\mathcal{D} \boldsymbol{u} = \frac{1}{2} \left( \nabla
+ \nabla^\top \right) \boldsymbol{u}$. The divergence of a tensor field is defined such that
$(\nabla \cdot \tensor{A})_j = \sum_{i=1}^{d} \partial A_{ij} / \partial x_i$.  Variables are
functions of position $\boldsymbol{x} \in \Omega$ and time $t \geq 0$.

\subsection{Governing partial differential equations}

We denote by $\rho \in \mathbb{R}$ the material density. The velocity field is introduced as
$\boldsymbol{u} (\boldsymbol{x}, t) \in \mathbb{R}^d$, with components $u$, $v$ and $w$.  The
Cauchy stress tensor $\tensor{\sigma}$ is defined as the sum of isotropic and deviatoric parts,
$\tensor{\sigma} = -p \tensor{I} + \tensor{\tau}$.  Here, the pressure $p (\boldsymbol{x}, t) \in
\mathbb{R}$ is multiplied by the identity tensor, while the deviatoric part of the stress tensor is
denoted $\tensor{\tau}(\boldsymbol{x},t) \in \mathbb{R}^{d \times d}_{\rm sym}$. The fluid motion
is then governed by incompressible mass conservation and momentum balance through the Cauchy
equation:
\begin{subequations}
\label{eq:governing}
\begin{align}
    \label{eq:incompressibility}
    \nabla \cdot \boldsymbol{u} &= 0 , \\
    \label{eq:momentum}
    \frac{\partial \boldsymbol{u}}{\partial t} + \boldsymbol{u} \cdot \nabla \boldsymbol{u}
    &= \frac{1}{\rho} \left( - \nabla p + \nabla \cdot \tensor{\tau} + \boldsymbol{f} \right) .
\end{align}
\end{subequations}
Here, we have introduced $\boldsymbol{f}$ to describe external body forces such as gravity acting
on the fluid.  Note that the mass conservation equation is simplified to
\eqref{eq:incompressibility} due to incompressibility, i.e.~constant density within a fluid parcel.

\subsection{Rheology}
%

We shall in the following restrict ourselves to relatively simple equations of state, of the form
$\tensor{\tau} = \tensor{\tau} (\dot{\tensor{\gamma}})$, where the stress response is solely
dependent on the rate-of-strain tensor $\dot{\tensor{\gamma}} = \mathcal{D} \boldsymbol{u} \in
\mathbb{R}^{n \times n}_{\rm sym}$. Although the Herschel-Bulkley model captures all aspects of the
fluids considered, references for validation are only available for power-law and Bingham fluids.
For this reason, we introduce these simpler rheological models first, and illustrate how they can be
combined.

Newtonian flow is characterised by a dynamic coefficient of viscosity $\mu > 0$ which is
independent of the strain, yielding a linear relationship between rate-of-strain and stress in the
rheological equation:
\begin{equation}
    \label{eq:rheology:newtonian}
    \tensor{\tau} = 2 \mu \dot{\tensor{\gamma}}.
\end{equation}
For fluids where the dependency of the stress on the rate-of-strain tensor is nonlinear, the
apparent viscosity is a useful concept when considering rheological responses to strain. It is a
generalisation of the constant viscosity for Newtonian flow, where we allow the viscosity to be a
function of the magnitude of the rate-of-strain tensor. Denoting the apparent viscosity by $\eta$,
we thus have
\begin{equation}
    \label{eq:apparent-viscosity}
    \eta = \frac{|\tensor{\tau}|}{2|\dot{\tensor{\gamma}}|} .
\end{equation}

Many fluids are accurately modelled by a non-Newtonian behaviour that captures shear-dependency
through a smooth increase or decrease in apparent viscosity. Such fluids include pseudoplastics
(shear-thinning, $\partial \eta / \partial |\dot{\tensor{\gamma}}| < 0$) and dilatants
(shear-thickening, $\partial \eta / \partial |\dot{\tensor{\gamma}}| > 0$).
A model which captures this behaviour is the power-law fluid, with rheological equation
\begin{equation}
    \label{eq:rheology:power-law}
    \tensor{\tau} = 2^{n} \mu |\dot{\tensor{\gamma}}|^{n-1} \dot{\tensor{\gamma}},
\end{equation}
and apparent viscosity
\begin{equation}
    \label{eq:viscosity:power-law}
    \eta = 2^{n-1} \mu |\dot{\tensor{\gamma}}|^{n-1} .
\end{equation}
A specific power-law fluid is characterised by its flow behaviour index $n > 0$ (and $\mu$, which
is usually referred to as the consistency for power-law fluids, and has units $\SI{}{Pa \, s}^n$).
From \eqref{eq:rheology:power-law} it is immediately clear that the Newtonian case with $n=1$
separates pseudoplastics ($n<1$) from dilatants ($n>1$).

Viscoplastic fluids have a stress threshold $\tau_0 > 0$ (the yield stress), below which they do
not flow. We note that it is possible to describe elastic deformation for materials which do not
flow, but we shall be considering constitutive equations which assume a rigid body approximation,
i.e.~zero strain rate below the yield stress.

The simplest type of viscoplastic fluid is the Bingham fluid \cite{bingham1916investigation,
oldroyd1947rational}, characterised by zero strain rate below the yield stress. In the yielded
region, however, the stress depends linearly on the rate-of-strain magnitude, just like a Newtonian
fluid. The stress-strain curve thus intercepts the $|\tensor{\tau}|$-axis at the point $(0,
\tau_0)$. As such, the Bingham fluid is a generalised Newtonian, with rheological equation
\begin{equation}
    \label{eq:rheology:bingham}
    \begin{cases}
        \dot{\tensor{\gamma}} = 0 &{\rm if} \quad |\tensor{\tau}| \leq \tau_0 \\ 
        \tensor{\tau} = 2 \mu \dot{\tensor{\gamma}} +
        \frac{\tensor{\tau_0}}{|\dot{\tensor{\gamma}}|} \dot{\tensor{\gamma}} &{\rm if} \quad
        |\tensor{\tau}| > \tau_0
    \end{cases}
\end{equation}
and apparent viscosity
\begin{equation}
    \label{eq:viscosity:bingham}
    \eta = \mu + \frac{\tau_0}{2 |\dot{\tensor{\gamma}}|} .
\end{equation}
For Bingham plastics, we refer to $\mu$ as the plastic viscosity, and note that $\eta$ has a
singularity for $\dot{\tensor{\gamma}}=0$, as expected.

\begin{figure*}
    \includegraphics[width=\textwidth]{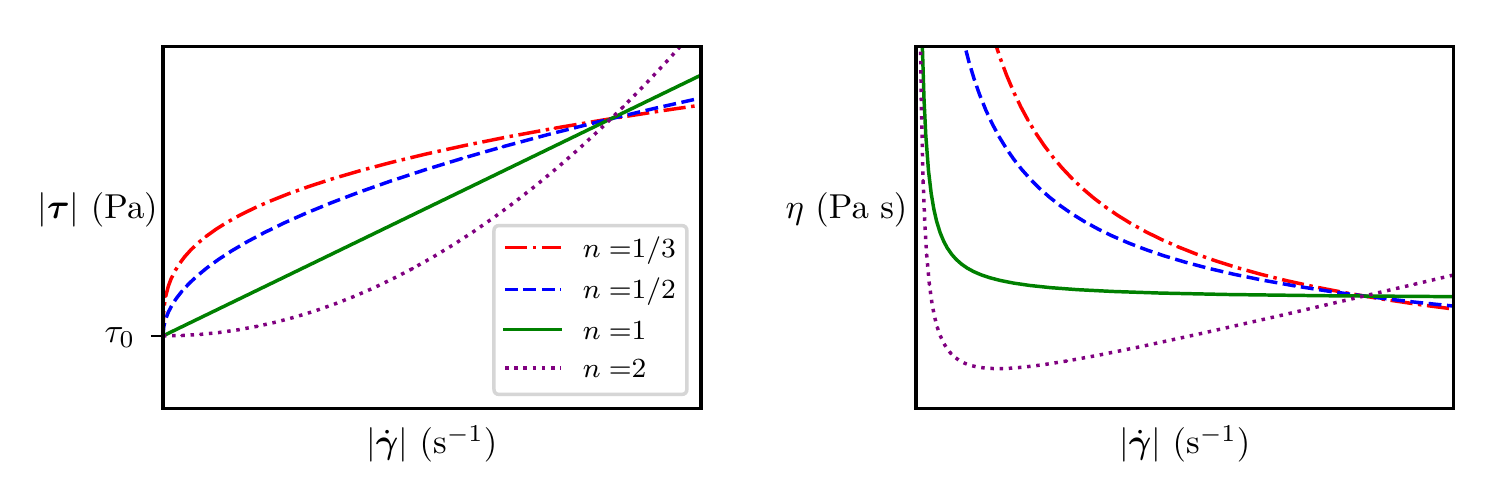}
    \caption
	{
		Herschel-Bulkley fluid: stress magnitude (left) and apparent viscosity (right) as functions
		of the magnitude of the rate-of-strain tensor $\dot{\gamma}$.
	}
    \label{fig:rheology:hb}
\end{figure*}

In many applications, it is desirable to capture both the yield stress of viscoplastic fluids and
the power-law dependency occurring once the fluid starts flowing. A widely used rheological model
for such fluids is due to Herschel and Bulkley \cite{herschel1926konsistenzmessungen}.
The Herschel-Bulkley fluid facilitates a very general description of non-Newtonian fluids,
as it is a yield stress fluid with a nonlinear stress-strain dependency in the
yielded region. As such, it can be thought of as a hybrid between Bingham plastics and power law
fluids. The constitutive equation is
\begin{equation}
    \label{eq:rheology:hb}
    \begin{cases}
        \dot{\tensor{\gamma}} = 0
        &{\rm if} \quad |\tensor{\tau}| \leq \tau_0 \\
        \tensor{\tau} = 2^{n} \mu |\dot{\tensor{\gamma}}|^{n-1} \dot{\tensor{\gamma}}
        + \frac{\tau_0}{|\dot{\tensor{\gamma}}|} \dot{\tensor{\gamma}}
        &{\rm if} \quad |\tensor{\tau}| > \tau_0
    \end{cases} ,
\end{equation}
while the apparent viscosity is
\begin{equation}
    \label{eq:viscosity:hb}
    \eta = 2^{n-1} \mu |\dot{\tensor{\gamma}}|^{n-1} +
    \frac{\tau_0}{2|\dot{\tensor{\gamma}}|}.
\end{equation}
Bingham plastics are a special case of Herschel-Bulkley fluids in exactly the same manner as
Newtonian fluids follow a specific power-law constitutive equation. Plots of the rheological
characteristic for Herschel-Bulkley fluids are shown in figure \ref{fig:rheology:hb}, for various
values of $n$. Note that the Bingham fluid is recovered for $n=1$.

\section{Numerical algorithm}
\label{sec:numerical}

\subsection{Incompressible flow solver}

An approximate projection method for solving the variable density incompressible Navier-Stokes
equations on an adaptive mesh hierarchy has been implemented in the
\href{https://github.com/AMReX-Codes/IAMR}{IAMR} code. The algorithm was described for the constant
viscosity case by Almgren et al.~ \cite{almgren1998conservative}, with extensions to low Mach
number reacting flows with temperature-dependent viscosity provided by Pember et
al.~\cite{pember1998adaptive} and Day and Bell \cite{day2000numerical}, among others.  In addition
to solving the Navier-Stokes equations for velocity and pressure, the IAMR code allows for the
(conservative or passive) advection of any number of scalar quantities.  The implementation is such
that the code can be run on architectures from single-core laptops through to massively parallel
supercomputers.

In the approximate projection method as implemented in IAMR, an advection-diffusion step is used to
advance the velocity in time; the solution is then (approximately) projected onto the space of
divergence-free fields. The motivation for an approximate rather than exact projection is covered
in detail in the literature \cite{lai1993projection, almgren1996numerical, martin2000cell}.

Velocity is defined at cell centres at integer time steps, and is denoted by
$\boldsymbol{u}_{i,j,k}^m$. The pressure, on the other hand, is specified at cell corners and is
staggered in time, so that it is denoted
$p_{i+\frac{1}{2},j+\frac{1}{2},k+\frac{1}{2}}^{m+\frac{1}{2}}$. When no subscript is given
(e.g.~$\boldsymbol{u}^m$), we address the distribution of $\boldsymbol{u}$ throughout the
computational domain at time $t^m$.  Considerations such as initialisation and boundary treatments
are ignored at present, and we set external forces equal to zero. Instead, we focus on the steps
required in order to evolve the solution a single time step $\Delta t$.

The first part of the advection-diffusion step is to extrapolate normal velocity components to
each cell face at time $t^{m+\frac{1}{2}}$. This is done through a second-order Taylor series
expansion. Consider the cell centre $(i,j,k)$, and suppose we want to extrapolate the first
velocity component $u$ to the face $(i+\frac{1}{2},j,k)$. Then the truncated Taylor series is
\begin{equation}
    u_{i+\frac{1}{2},j,k}^{m+\frac{1}{2}} \approx
    u_{i,j,k}^m + \frac{\Delta x}{2} \left( \frac{\partial u}{\partial x} \right) _{i,j,k}^m +
    \frac{\Delta t}{2} \left( \frac{\partial u}{\partial t} \right)_{i,j,k}^m.
\end{equation}
Taking the first component of \eqref{eq:momentum}, we have
\begin{multline}
    \left( \frac{\partial u}{\partial t} \right)_{i,j,k}^m =
    - \left(u \frac{\partial u}{\partial x} \right)_{i,j,k}^m
    - \left(v \frac{\partial u}{\partial y} \right)_{i,j,k}^m \\
    - \left(w \frac{\partial u}{\partial z} \right)_{i,j,k}^m 
    - \frac{1}{\rho}
    \left( \frac{\partial p}{\partial x} \right)_{i,j,k}^{m-\frac{1}{2}} \\
    + \frac{1}{\rho} \left(
    \frac{\partial \tau_{11}}{\partial x}
    + \frac{\partial \tau_{21}}{\partial y}
    + \frac{\partial \tau_{31}}{\partial z}
    + \right)_{i,j,k}^{m} ,
\end{multline}
which can be substituted into the last term of the Taylor expansion to avoid any dependence on
temporal derivatives. In order to compute these extrapolated values, however, it is necessary to
calculate discrete approximations of the spatial derivatives. This is done separately for each
part. The first derivative of normal velocity is evaluated using a monotonicity-limited
fourth-order slope approximation $S$ as introduced by Colella in 1985
\cite{colella1985direct}. Since the pressure field is defined at grid nodes, cell-centred
approximations of the pressure gradient $\G$ are computed by finite differences.
Finally, transverse derivative terms such as $v \partial u / \partial y$ are evaluated by a
separate extrapolation and subsequent upwinding procedure,
\begin{equation}
    \left( v \frac{\partial u}{\partial y} \right)_{i,j,k}^m
    \approx \left( \widehat{v \frac{\partial u}{\partial y}} \right)_{i,j,k}^m.
\end{equation}
With
these discrete representations in place, the intermediate normal velocity at face $i+\frac{1}{2}$
can be approximated based on the cell to its left ($i,j,k$) as
\begin{subequations}
\label{eq:iamr:extrap}
\begin{multline}
    \label{eq:iamr:extrap:left}
    u_{i+\frac{1}{2},j,k}^{m+\frac{1}{2}, L} =
    u_{i,j,k}^m + \left( \frac{\Delta x}{2} - u_{i,j,k}^m \frac{\Delta t}{2} \right)
    (S u)_{i,j,k}^m \\
    - \frac{\Delta t}{2}
    \left( \widehat{v \frac{\partial u}{\partial y}} \right)_{i,j,k}^m
    - \frac{\Delta t}{2}  \left( \widehat{w \frac{\partial u}{\partial z}} \right)_{i,j,k}^m \\
    - \frac{\Delta t}{2 \rho} \left(
    \left( -(\G_x p)_{i,j,k}^{m-\frac{1}{2}} +
    \left( \D_x \tensor{\tau} \right)_{i,j,k}^m \right) \right).
\end{multline}
Similarly, an approximation based on the state to its right ($i+1,j,k$) is
\begin{multline}
    \label{eq:iamr:extrap:right}
    u_{i+\frac{1}{2},j,k}^{m+\frac{1}{2}, R} =
    u_{i+1,j,k}^m + \left( \frac{\Delta x}{2} + u_{i+1,j,k}^m \frac{\Delta t}{2} \right)
    (S u)_{i+1,j,k}^m \\
    - \frac{\Delta t}{2}
    \left( \widehat{v \frac{\partial u}{\partial y}} \right)_{i+1,j,k}^m
    - \frac{\Delta t}{2} \left( \widehat{w \frac{\partial u}{\partial z}} \right)_{i+1,j,k}^m \\
    - \frac{\Delta t}{2 \rho}
    \left( -(\G_x p)_{i+1,j,k}^{m-\frac{1}{2}} + \left( \D_x \tensor{\tau}
    \right)_{i+1,j,k}^m \right).
\end{multline}
\end{subequations}
These extrapolations need to be computed for tangential velocity components as well.  At each face
we then have two extrapolated approximations for $\boldsymbol{u}^{m+\frac{1}{2}}$, one from each
cell adjoining at the face. In order to pick the most accurate of these, we construct a face-based
advection velocity $\boldsymbol{u}^{\rm adv}$.  The final, upwinded, extrapolated approximation at
each face is denoted $\tilde{\boldsymbol{u}}^{m+\frac{1}{2}}$.

With these extrapolated and upwinded approximations in place, \eqref{eq:momentum} is discretised in
time to construct a new-time provisional velocity field, $\bu^*$, without enforcing
\eqref{eq:incompressibility}, i.e.~we define $\bu^{\ast}$ using
\begin{multline}
  \frac{\bu^{\ast} - \bu^m}{\dt} = - [\bu^{\rm MAC} \cdot \nabla \tilde{\bu}]^\nph \\
    + \frac{1}{\rho} \left( - {\G} p^\nmh + \frac{1}{2}
	\left( \nabla \cdot \tensor{\tau}(\bu^m)  +  \nabla \cdot \tensor{\tau}(\bu^\ast)  \right)
	+ \boldsymbol{f}^m \right) ,
\end{multline}
where $\G p^{\nmh}$ is a lagged approximation to the pressure gradient $\nabla p$ and the density
$\rho$ is a constant\footnote{IAMR is designed for more general flows with variable density;
however for the flows considered in this paper the density is constant in space and time.}.

The explicit viscous term, $\tensor{\tau}(\bu^m),$ is evaluated using only the velocity components
at time $t^m$, i.e.~we define $\dot{\tensor{\gamma}}^m = \dot{\tensor{\gamma}}(\bu^m)$, $\eta^m =
\eta(\dot{\tensor{\gamma}}^m)$ and write $\tensor{\tau}(\bu^m)  = 2 \eta^m \;
\dot{\tensor{\gamma}}^m$.  The implicit term, $\tensor{\tau}(\bu^\ast)$ as well as $\bu^\ast$
itself, are solved for with the same $\eta^m$,  through a tensor solve of the form
\begin{multline}
    \left(
    \bu^{\ast} - \frac{\dt}{\rho} \nabla \cdot (\eta^m \;  \dot{\tensor{\gamma}}^\ast)
    \right)
    = \bu^m - \dt [\bu^{\rm MAC} \cdot \nabla \tilde{\bu}]^\nph \\
	+ \frac{\dt}{\rho}
	\left( - \G p^\nmh + \nabla \cdot (\eta^m {\tensor{\gamma}^m}) + \boldsymbol{f}^m \right) .
\end{multline}
Note that all velocity components are solved for simultaneously.

The final part of the algorithm is the projection step and subsequent pressure update. The velocity
field $\boldsymbol{u}^*$ does not in general satisfy the divergence constraint as given by
\eqref{eq:incompressibility}.   We solve
\begin{equation}
    L_{{\rho}} \phi = D \left( \frac{1}{\dt} \; \bu^\ast +  \frac{1}{\rho} \; \G p^\nmh \right)
\end{equation}
where $D$ is a discrete divergence and $\G$ a discrete gradient. $L_\rho$ is a second-order
accurate approximation to $\nabla \cdot \frac{1}{\rho} \nabla.$
The new-time velocity is then defined by
\begin{equation}
    \bu^{m+1} = \bu^\ast - \dt \frac{1}{\rho} \G \phi
\end{equation}
and the updated pressure by
\begin{equation}
    p^{\nph} = \phi \enskip .
\end{equation}
The resulting approximate projection satisfies the divergence constraint to second-order accuracy
and ensures stability for the algorithm. Given suitable initial and boundary conditions, this
algorithm can be used to evolve the system in time. For time-dependent flows approaching a
steady-state solution, we consider the system steady when
\begin{equation}
    \max \left| \frac{\bu^{m+1} - \bu^m}{\bu^m} \right| < 10^{-5} .
\end{equation}
For further detail on the algorithm, see the original paper by Almgren et
al.\cite{almgren1998conservative}

\subsection{Regularisation of viscosity}

Equation \eqref{eq:rheology:hb} does not present any problems for unyielded regions
(${\dot{\tensor{\gamma}} > 0}$), but the apparent viscosity has a singularity when the strain rate
approaches zero (and ${|\tensor{\tau}| \rightarrow \tau_0}$). Computational schemes such as those
implemented in IAMR cannot be used in the presence of such singularities. Regularisation deals with
the problem by replacing the ill-behaved apparent viscosity with a function that approximates the
rheological behaviour, but which stays bounded for arbitrarily small $\dot{\tensor{\gamma}}$. This
is done by introducing an additional parameter $\varepsilon$ to the apparent viscosity, which
describes how big the effect of the regularisation is. A large value of $\varepsilon$ allows for
inexpensive computations even near unyielded flow, while the limit $\varepsilon \rightarrow 0$
recovers the unregularised description. Note that one must be careful to choose a small enough
value for $\varepsilon$, in addition to fine enough mesh resolutions and convergence criterion.
Otherwise, the mathematical approximation will not hold.

\begin{figure*}
    \includegraphics[width=\textwidth]{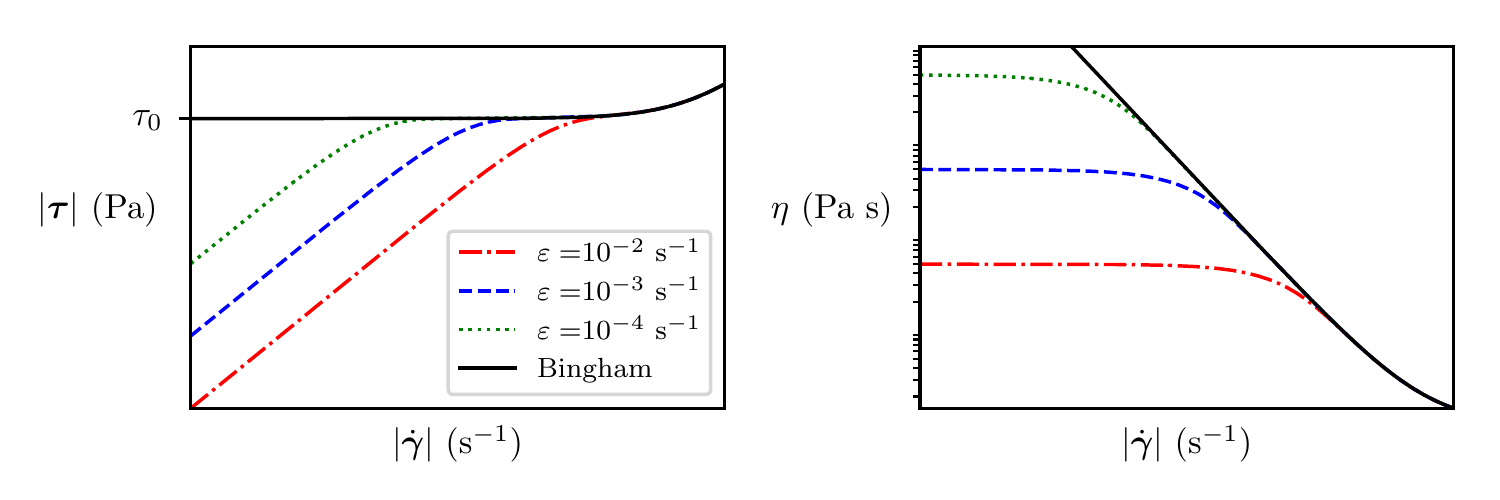}
	\caption
    {
        Papanastasiou regularisation for the Bingham model: stress magnitude (left) and apparent
        viscosity (right) as functions of the magnitude of the rate-of-strain tensor
        $|\dot{\tensor{\gamma}}|$. For decreasing $\varepsilon$, we recover a closer
        approximation to the actual Bingham model. Note that logarithmic axes are employed in order
        to highlight the behaviour in the low-strain limit.
    }
    \label{fig:rheology:papanastasiou}
\end{figure*}

We employ the popular Papanastasiou regularisation
\cite{papanastasiou1987flows,mirzaagha2017rising}, which utilises an
exponential relaxation according to
\begin{equation}
    \label{eq:regularisation}
	\frac{1}{|\dot{\tensor{\gamma}}|} \rightarrow
	\frac{1 - e^{-|\dot{\tensor{\gamma}}| / \varepsilon}}{|\dot{\tensor{\gamma}}|}.
\end{equation}
This is a good approximation for ${|\dot{\tensor{\gamma}}| / \varepsilon \gg 1}$, while in the
small strain limit we have
\begin{equation}
  \label{eq:regularisation:limit}
  \lim_{|\dot{\tensor{\gamma}}| \rightarrow 0}
  \frac{1 - e^{-|\dot{\tensor{\gamma}}| / \varepsilon}}{|\dot{\tensor{\gamma}}|} =
  \lim_{|\dot{\tensor{\gamma}}| \rightarrow 0}
  \frac{1}{|\dot{\tensor{\gamma}}|}
  \left( 1 - \sum_{k=0}^{\infty} \frac{(-|\dot{\tensor{\gamma}}| / \varepsilon)^k}{k!} \right) =
  \frac{1}{\varepsilon} ,
\end{equation}
so that it always remains bounded, and recovers the unregularised model in the limit $\varepsilon
\rightarrow 0$. Combination of \eqref{eq:regularisation} with the apparent viscosity as given in
\eqref{eq:rheology:hb}, gives the regularised viscosity as
\begin{equation}
    \label{eq:rheology:papanastasiou}
	\eta =
	\left( 2^{n-1} \mu |\dot{\tensor{\gamma}}|^{n-1} +
	\frac{\tau_0}{2|\dot{\tensor{\gamma}}|} \right)
    \left( 1 - e^{-|\dot{\tensor{\gamma}}| / \varepsilon} \right) .
\end{equation}
The effect of varying the regularisation parameter $\varepsilon$ is shown in figure
\ref{fig:rheology:papanastasiou}. In the Papanastasiou-regularised Herschel-Bulkley model, the
singularity in apparent viscosity is replaced by the limiting value
\begin{equation}
  \lim_{|\dot{\tensor{\gamma}}| \rightarrow 0} \eta = \frac{\tau_0}{2 \varepsilon}.
\end{equation}

\subsection{Adaptive mesh refinement}

For a vast range of flow problems, large computational domains are required even though we are only
interested in small areas of the domain such as boundaries, interfaces or areas where variables
exhibit complex dynamic behaviour. Refining the computational mesh in these regions is a well-known
technique in order to save memory and processing power, and it can be applied with various degrees
of sophistication. Within \href{https://amrex-codes.github.io/IAMR/}{IAMR}, and the
\href{https://amrex-codes.github.io/amrex/}{AMReX} framework in general, hierarchical structured
AMR is employed, which allows for efficient local refinement by symmetrically splitting each cell
into $2^d$ smaller ones, where $d$ is the dimensionality of the problem. Furthermore, the dynamic
refinement is spatio-temporal, meaning that we employ subcycling in the discrete time steps to
avoid setting global time step restrictions based on the smallest cells. This is crucial for dynamic
AMR to be used efficiently. Tagging criteria are utilised in order for the code to determine which
cells should be split or merged at a given refinement step, and these can depend on the values of
any directly simulated or derived variables, in addition to spatial position and time. 

For details on the implementation of structured AMR in this code, we refer to the websites of AMReX
(\href{https://amrex-codes.github.io/amrex/}{https://amrex-codes.github.io/amrex/}) and IAMR
(\href{https://amrex-codes.github.io/IAMR/}{https://amrex-codes.github.io/IAMR/}), where full
documentation and source code is available.

\subsection{Parallel implementation}

\href{https://github.com/AMReX-Codes/amrex}{AMReX} is a mature, open-source software framework for
building massively parallel block-structured AMR applications.  AMReX contains extensive software
support for explicit and implicit grid-based operations. Multigrid solvers, including those for
tensor systems, are included for cell-based and node-based data.  AMReX uses a hybrid MPI/OpenMP
approach for parallelization; in this model individual grids are distributed to MPI ranks, and
OpenMP is used to thread over logical tiles within the grids.  Applications based on AMReX have
demonstrated excellent strong and weak scaling up to hundreds of thousands of cores
\cite{almgren2010castro, almgren2013nyx, dubey2014survey, habib2016hacc}.


\section{Validation}
\label{sec:validation}

In order to validate our code, we consider the two-dimensional lid-driven cavity test case. It is a
historically significant and widely used benchmark problem for viscous flow simulations, and
consequently a large amount of reference results exist in the literature. Our domain is a square of
side length $\mathcal{L}$, i.e.~$\Omega = [0,\mathcal{L}]^2$, and is filled with a fluid of
constant density. Initially, the system is at rest. All walls except the top (the ``lid'') are held
fixed, so that the boundary conditions on these walls are $\boldsymbol{u} = (0, 0)$. At time $t=0$,
the lid is instantaneously prescribed the tangential velocity $\mathcal{U}$, so that the relevant
boundary condition for $t>0$ is $\boldsymbol{u} = (\mathcal{U}, 0)$. Without any other external
forces ($\boldsymbol{f} = 0$), this drives a recirculating flow in the cavity, which reaches a
steady-state solution for non-turbulent regimes.

In order to obtain a dimensionless form of \eqref{eq:governing}, we let ${\hat{\boldsymbol{x}} =
\boldsymbol{x} / \mathcal{L}}$, ${\hat{\boldsymbol{u}} = \boldsymbol{u} / \mathcal{U}}$ and
${\hat{t} = t / (\mathcal{L} / \mathcal{U})}$. Additionally, we take $\mathcal{U} /
\mathcal{L}$ as a characteristic strain-rate, and let
\begin{align}
    \hat{\dot{\tensor{\gamma}}} &= \frac{\dot{\tensor{\gamma}}}{\frac{\mathcal{U}}{\mathcal{L}}},
    &\quad
    \hat{\eta} &= \frac{\eta}{2^{n-1} \mu
    \left( \frac{\mathcal{U}}{\mathcal{L}} \right)^{n-1}}, \\
    \hat{p} &= \frac{p}{2^{n-1} \mu
    \left( \frac{\mathcal{U}}{\mathcal{L}} \right)^n}, &\quad
    \hat{\tensor{\tau}} &= \frac{\tensor{\tau}}{2^{n-1} \mu
    \left( \frac{\mathcal{U}}{\mathcal{L}} \right)^n}.
\end{align}
By substituting these dimensionless variables into \eqref{eq:governing}, we find that the
governing equations in dimensionless form are
\begin{subequations}
\begin{align}
    \label{eq:dimensionless:momentum}
    \frac{\partial \hat{\boldsymbol{u}}}{\partial \hat{t}} + \hat{\boldsymbol{u}} \cdot
    \hat{\nabla} \hat{\boldsymbol{u}}
    &= \frac{1}{\Rey}
    \left( - \hat{\nabla} \hat{p} + \hat{\nabla} \cdot \hat{\tensor{\tau}} \right) , \\
    \label{eq:dimensionless:incompressibility}
    \hat{\nabla} \cdot \hat{\bu} &= 0 , \\
    \label{eq:dimensionless:stress}
    \hat{\tensor{\tau}} &= 2 \hat{\eta} \hat{\dot{\tensor{\gamma}}} , \\
    \label{eq:dimensionless:herschel-bulkley}
    \hat{\eta} &= \left(|\hat{\dot{\tensor{\gamma}}}|^n + \frac{1}{2} \HB\right)
    \frac{1 - e^{-\Pap |\hat{\dot{\tensor{\gamma}}}|}}{|\hat{\dot{\tensor{\gamma}}}|} ,
\end{align}
\end{subequations}
where we have introduced the dimensionless groups
\begin{equation}
    \Rey = \frac{\rho \mathcal{U}^2}{2^{n-1}
    \mu \left(\frac{\mathcal{U}}{\mathcal{L}}\right)^n} , \quad
    \HB = \frac{\tau_0}{2^{n-1} \mu \left(\frac{\mathcal{U}}{\mathcal{L}}\right)^n} ,
    \quad
    \Pap = \frac{\mathcal{U}}{\varepsilon \mathcal{L}} .
\end{equation}

The Reynolds number $\Rey$ is the ratio of inertial forces to (power-law) viscous ones, while the
Herschel-Bulkley number $\HB$ quantifies the effect of yield stress versus power-law viscosity. The
Papanastasiou number $\Pap$ measures the degree of regularisation employed in the apparent
viscosity, with higher $\Pap$ being a more accurate description of unregularised viscoplasticity.
It is worth noting that the Reynolds number for Newtonian flow ($n=1$) is $\Rey_N = \rho
\mathcal{U} \mathcal{L} / \mu$.

When we have information about the dimensionless apparent viscosity $\hat{\eta}$ throughout our
domain at a given time, we can compute the local Reynolds number field $Re_L = Re / \hat{\eta}$.
This value, which is proportional to the inverse of the apparent viscosity, provides insight about
which regions of the domain are dominated by the effects of viscoplasticity, and which ones have
near-Newtonian behaviour.  Finally, for the special case $n=1$, we introduce the Bingham number
$\Bin = \tau_0 \mathcal{L} / \mu \mathcal{U}$, to allow comparisons with articles written on
simulation of Bingham fluids.

In the remainder of this section, we use $\mathcal{L} = \SI{1}{m}$, $\rho = \SI{1000}{kg/m^3}$,
$\mathcal{U} = \SI{1}{m/s}$ and $\Pap = 400$. It is important to choose a high enough value of
$\Pap$ when using regularised schemes, and our choice is based on the arguments by Syrakos et
al.~\cite{syrakos2013solution}, who show that this value results in highly accurate simulations.
The cavity is discretised spatially with 256 cells in each dimension.  The lid-driven cavity flow
is then uniquely defined by the choices of $\mu$, $n$ and $\tau_0$.

To the authors' best knowledge, reference results for Herschel-Bulkley fluids have not been
presented in the literature. There are, however, plenty of results for power-law fluids and Bingham
plastics. Since these two characterise separate components of the Herschel-Bulkley model, we expect
that validating each of them individually is a good test for our code. Evaluation of the full
Herschel-Bulkley model, based on this validation, is given in section \ref{sec:evaluation}.

Power-law fluids have zero yield stress, and are different from the Newtonian case when the flow
behaviour index is different from unity. They were studied using a least squares finite elements
formulation by Bell and Surana in 1994 \cite{bell1994p}, who presented results with $n=1/2$ and
$n=3/2$. A third-order finite volume method with upwinding was then used to study the same problem
by Neofytou in 2005 \cite{neofytou20053rd}. The classical way to compare results from the
lid-driven cavity test case is to drive the system to steady state, and then look at the velocity
profiles in each direction at slices perpendicular to the flow. This ensures that we can compare
with results from codes that can only calculate steady-state solutions.

\begin{figure*}
  \includegraphics[width=\textwidth]{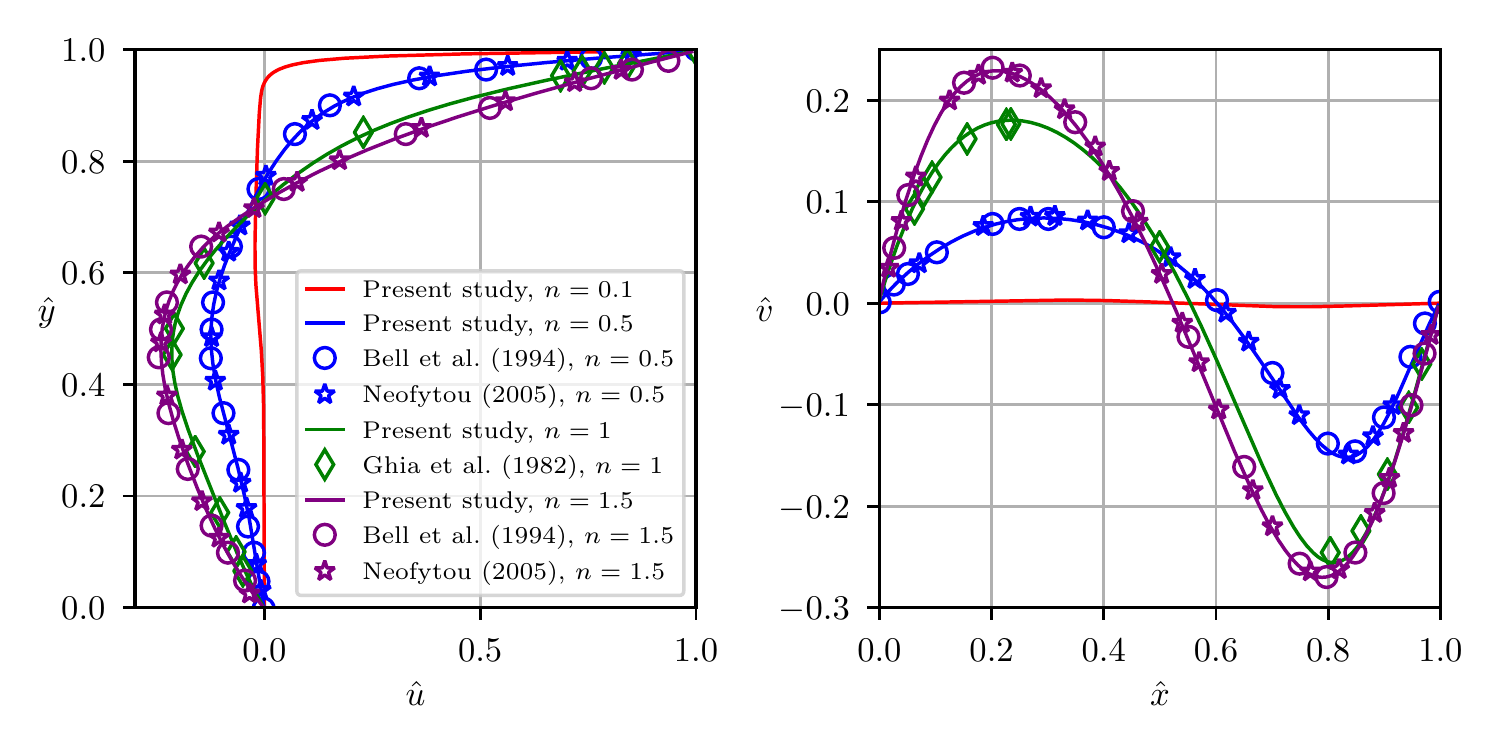}
  \caption
  {
      Velocity profiles for power-law fluids with $\mu = \SI{10}{Pa \, s}^n$. Left: First
      component of velocity through the vertical slice $\hat{x}=0.5$. Right: Second component of
      velocity through the horizontal slice $\hat{y}=0.5$.
  }
  \label{fig:velocity_profiles_powerlaw}
\end{figure*}

Figure \ref{fig:velocity_profiles_powerlaw} shows our results for power-law fluid velocity
profiles, and includes reference results from the aforementioned papers. Additionally, we include
the Newtonian case with comparisons from the classical paper by Ghia et al.~\cite{ghia1982high}. It
is evident that decreasing $n$ results in less lid-induced kinetic energy propagating further down
in the domain, whereas increasing it has the opposite effect. Our results align very well with
those found in the literature, especially those due to Neofytou, which have the highest accuracy.
We have also included the velocity profile for $n=0.1$, in order to illustrate the significant
viscous resistance in the limit $n \rightarrow 0$.

\begin{figure*}
  \includegraphics[width=\textwidth]{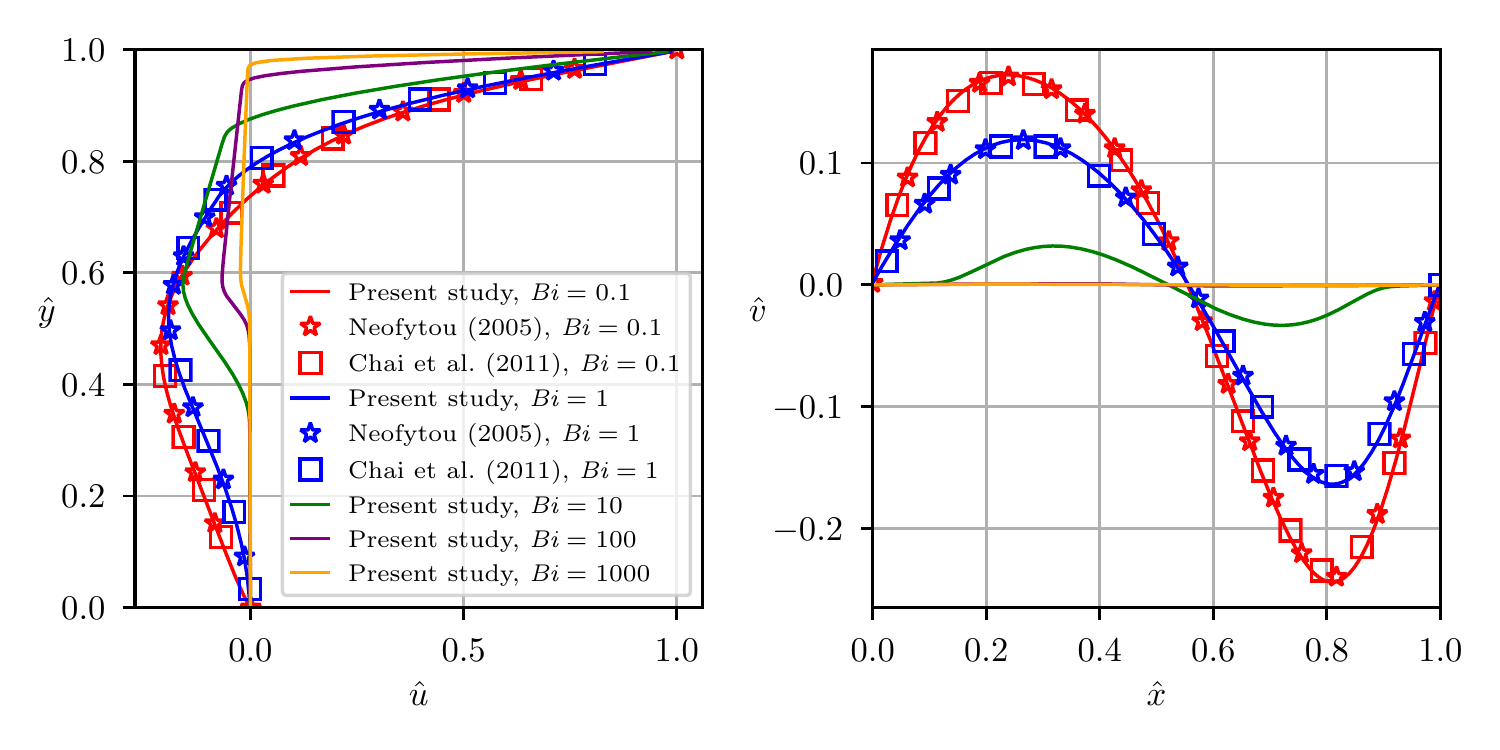}
  \caption
  {
      Velocity profiles for Bingham fluids with $\Rey=100$. Left: First component of velocity
      through the vertical slice $\hat{x}=0.5$. Right: Second component of velocity through the
      horizontal slice $\hat{y}=0.5$.
  }
  \label{fig:velocity_profiles_bingham_Re100}
\end{figure*}

We recall that Bingham fluids have $n=1$ and non-zero yield stress $\tau_0$. Due to the importance
of accurately capturing viscoplastic behaviour, we have performed extra validation tests for this
model. Figure \ref{fig:velocity_profiles_bingham_Re100} depicts velocity profiles at various
Bingham numbers, with comparison to the works of Neofytou \cite{neofytou20053rd} and Chai et
al.~\cite{chai2011multiple}. In the latter reference, a multiple-relaxation-time lattice Boltzmann
model was employed to simulate non-Newtonian fluids. We see that increasing the yield-stress has an
effect on the velocity profiles similar to that of lowering $n$ for the power-law fluids, although
the Bingham fluids have sharper transitions in the profile (especially noteworthy in the
$x$-velocity).  Our results are in excellent agreement with the relevant references, and we have
also extended the range of Bingham numbers beyond those for which velocity profiles are available
in the literature.

\begin{figure}
	\includegraphics[width=\columnwidth]{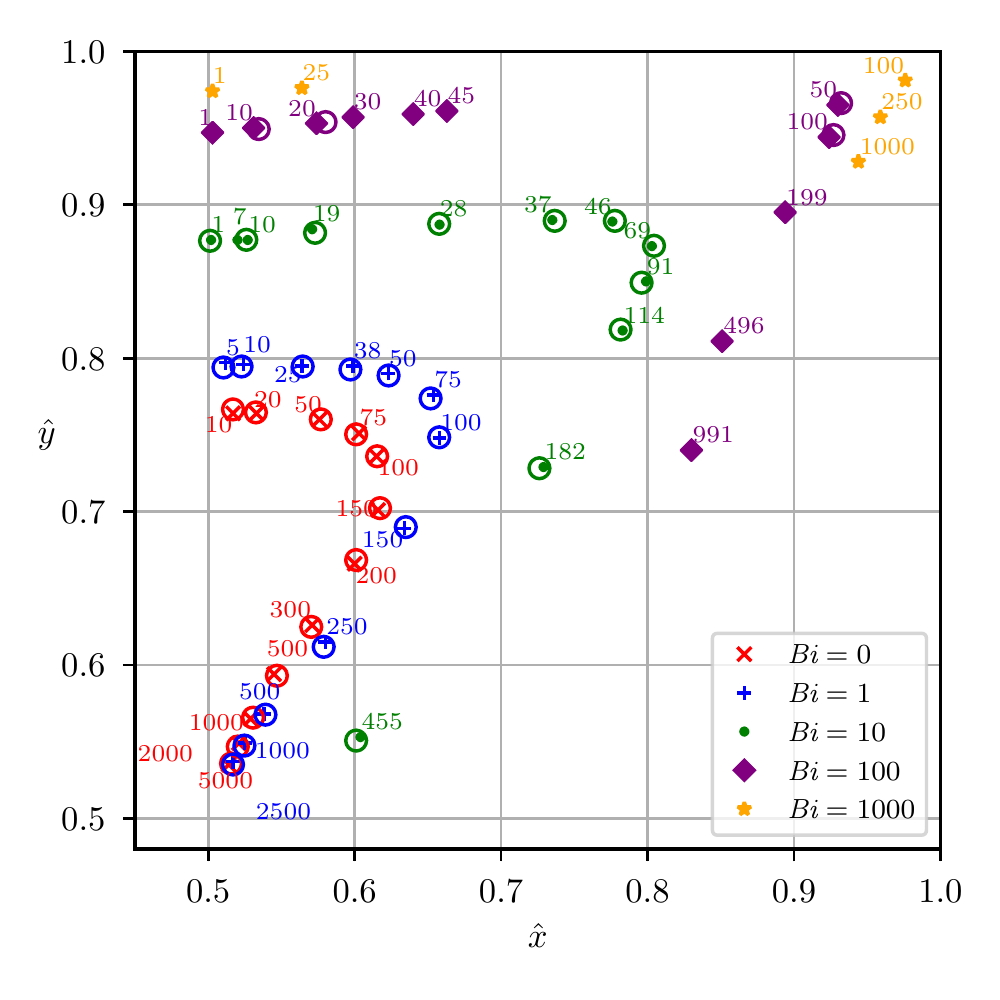}
	\caption
	{
        Vortex centre locations in the two-dimensional lid-driven cavity for Bingham fluids.
        Empty circles are results from Syrakos et al.~\cite{syrakos2014performance}, while the
        remaining symbols are results from the present study. The number next to each point is the
        corresponding modified Reynolds number $\Rey^*$.
	}
	\label{fig:vortex_centers}
\end{figure}

The second method we use for validation is locating the position of the main vortex centre within
the cavity, at steady-state. A comprehensive study exploring this dependency was performed in 2014
by Syrakos et al.~\cite{syrakos2014performance}. We have performed simulations to steady-state for
a large number of configurations of $\Rey$ and $\Bin$, and the resulting vortex locations are
depicted in figure \ref{fig:vortex_centers}, alongside the results of Syrakos et al. We reiterate
what they found in their paper: increasing the Bingham number moves the vortex upwards and to the
right, while increasing the Reynolds number moves the vortex first towards the right, and then
downwards and left to the centre. Our results agree very well with the reference results for the
range of ($\Rey$, $\Bin$) pairs covered in that study. Additionally, we have covered many more
large Bingham numbers, obtaining results which follow the patterns to be expected. This illustrates
the capability of our code to simulate fluids with very large yield stress. Please note that
the number next to each point in the figure is the modified Reynolds number
\begin{equation}
    \label{eq:reynolds_modified}
    \Rey^* = \frac{\Rey}{(\Bin + 1)} .
\end{equation}
This is because, following Syrakos et al.~\cite{syrakos2016cessation}, it provides a more natural
measure of the transition to turbulence in the high-$\Bin$ region.  For further discussions on
choice of dimensionless parameters in viscoplastic fluid mechanics, we refer the reader to articles
by de Souza Mendes \cite{mendes2007dimensionless} and Thompson and Soares
\cite{thompson2016viscoplastic}.

The most important validation for viscoplastic fluids, is the accurate determination of yield
surfaces. Regions where the stress satisfies $|\tensor{\tau}| \leq \tau_0$ are known as unyielded
regions. The interface which separates yielded and unyielded regions is called the yield surface.
Since the yield surface separates two fundamentally different states of matter (viscous fluid flow
and rigid solid behaviour), locating it precisely is a valuable metric for evaluating numerical
results. Finally, we would like to ensure that our time-marching scheme accurately captures the
fluid movement. In order to validate both of these two attributes, we run a simulation starting
from rest and evolving to the steady-state for the lid-driven cavity. Note that the criteria for
the system having reached steady-state is that the maximum relative change in the velocity field
from one iteration to the next is less than some tolerance (in our case $10^{-5}$). At this point,
we set the lid velocity to zero, and let the system come to rest again through cessation. When the
entire domain is an unyielded region, we stop the simulation. Although some authors have used
time-stepping schemes to advance the system to steady-state
\cite{dean2007numerical,muravleva2015uzawa}, we are not aware of any results illustrating the yield
surface development for instantaneous start from rest. On the other hand, the latter half of this
numerical experiment, i.e.~cessation of Bingham fluids from the lid-driven steady-state, was
studied numerically by Syrakos et al.~in 2016 \cite{syrakos2016cessation}.  Comparisons with their
yield surfaces and time measurements therefore serve as validation for our code.

\begin{figure*}
	\includegraphics[width=0.8\textwidth]{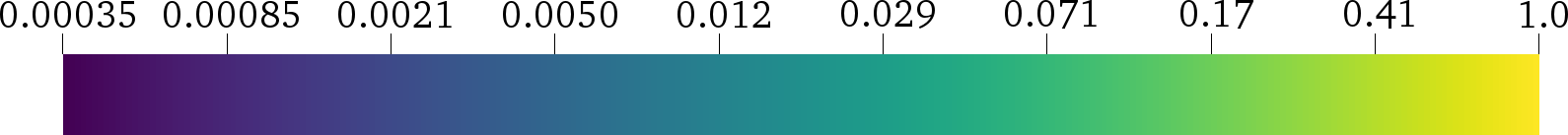}
	\\[1em]
	\begin{subfigure}{0.35\textwidth}
		\includegraphics[width=\textwidth]{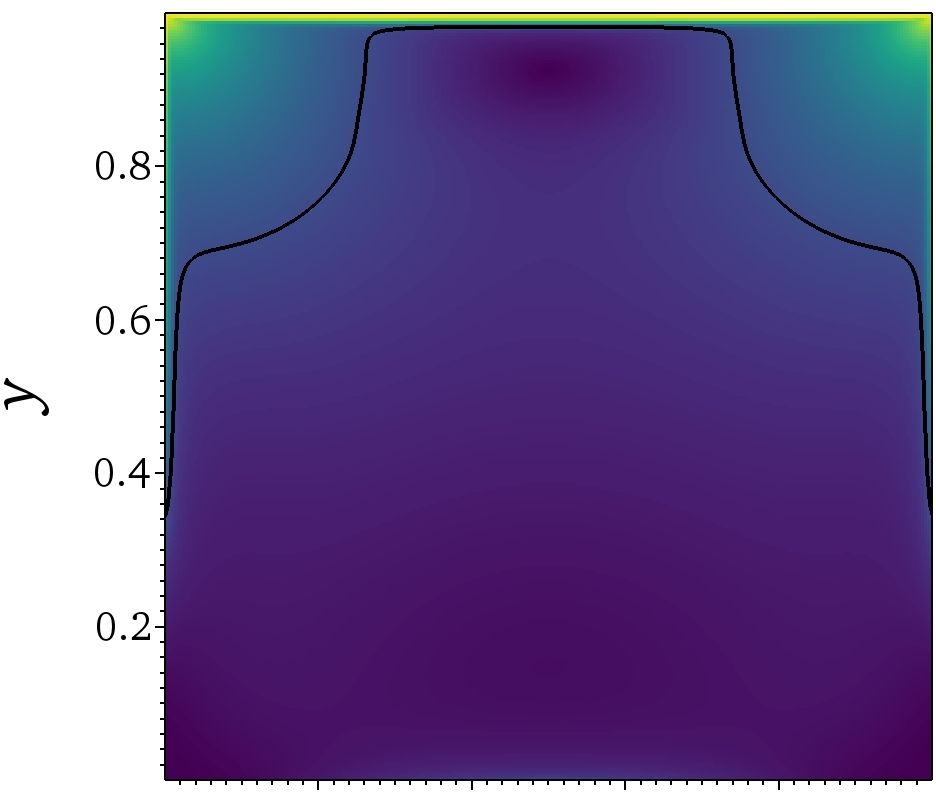}
		\caption{$\hat{t}=3.5 \times 10^{-6}$}
	\end{subfigure}
	~
	\begin{subfigure}{0.29\textwidth}
		\includegraphics[width=\textwidth]{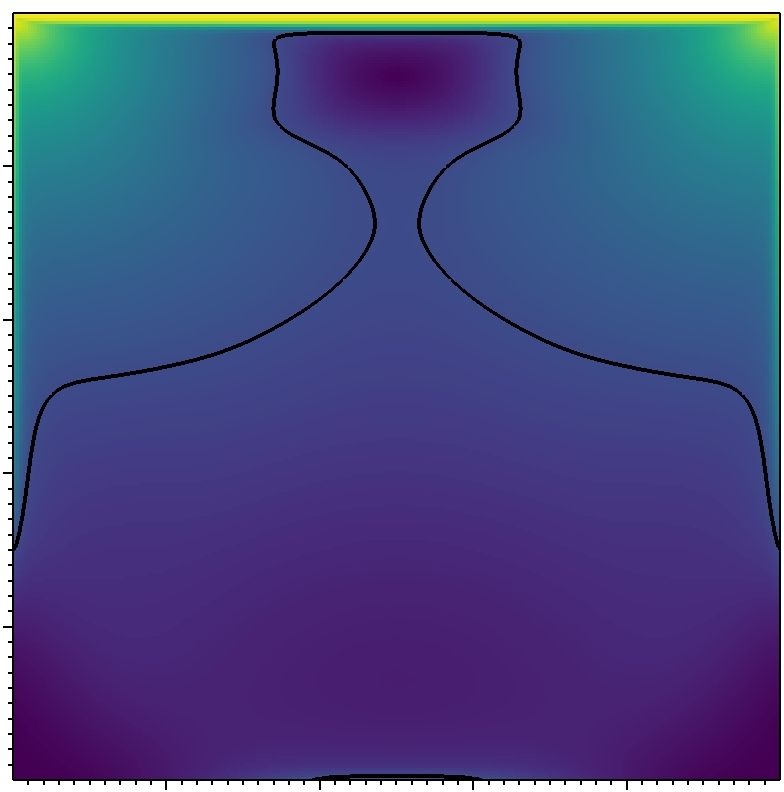}
		\caption{$\hat{t}=1.05 \times 10^{-5}$}
	\end{subfigure}
	~
	\begin{subfigure}{0.29\textwidth}
		\includegraphics[width=\textwidth]{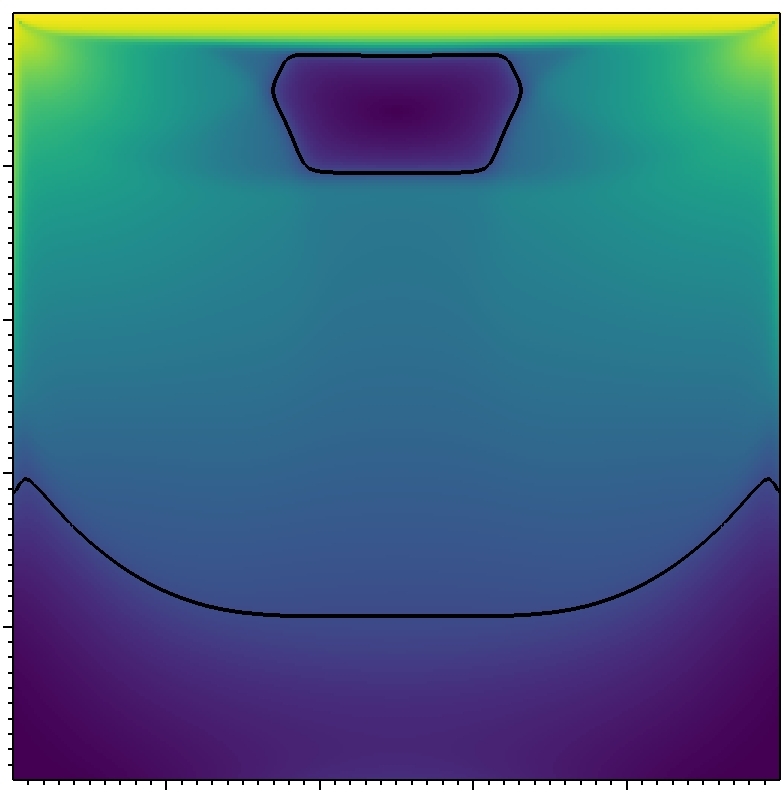}
		\caption{$\hat{t}=9.33 \times 10^{-5}$}
	\end{subfigure}
	\\[1em]
	\begin{subfigure}{0.35\textwidth}
		\includegraphics[width=\textwidth]{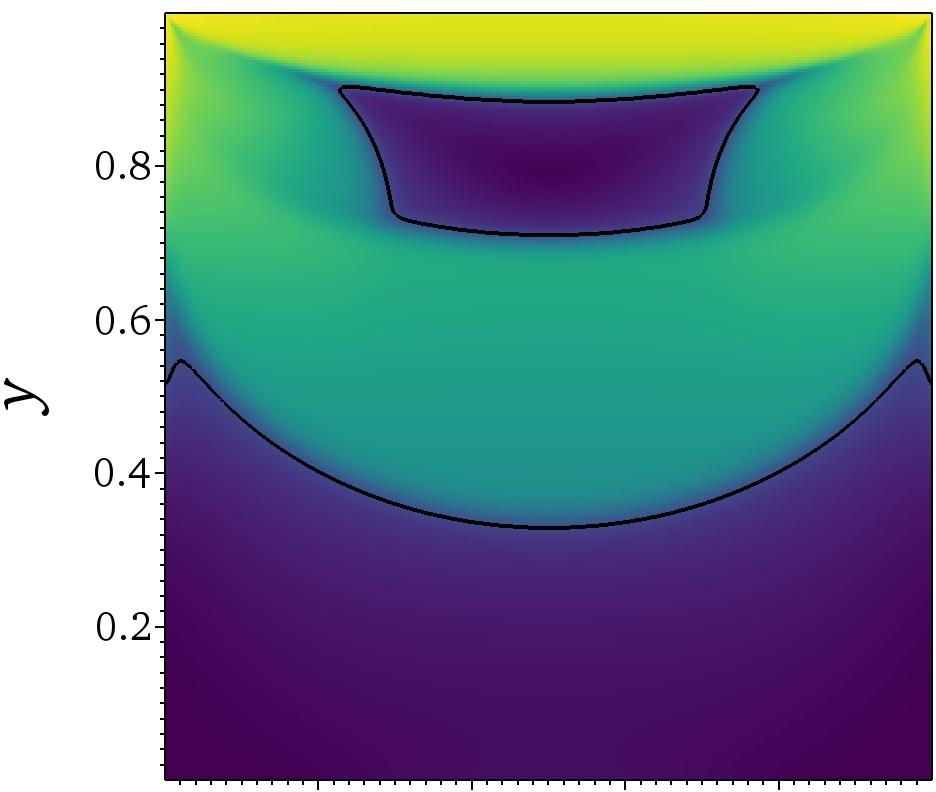}
		\caption{$\hat{t}=9.540 \times 10^{-4}$}
	\end{subfigure}
	~
	\begin{subfigure}{0.29\textwidth}
		\includegraphics[width=\textwidth]{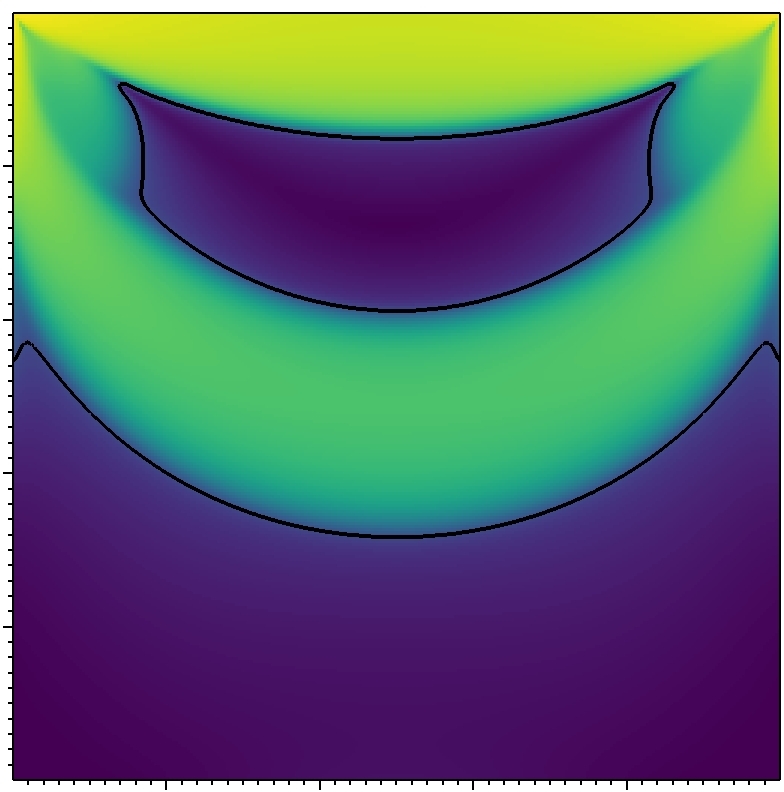}
		\caption{$\hat{t}=0.020555, \quad \tilde{t}=0$}
	\end{subfigure}
	~
	\begin{subfigure}{0.29\textwidth}
		\includegraphics[width=\textwidth]{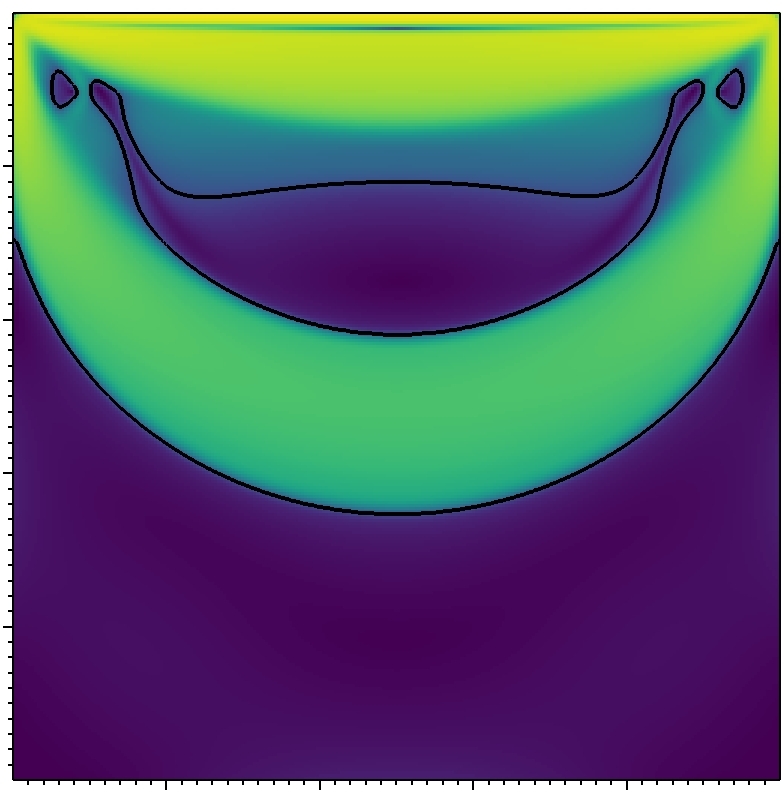}
		\caption{$\tilde{t}=3.89 \times 10^{-5}$}
	\end{subfigure}
	\\[1em]
	\begin{subfigure}{0.35\textwidth}
		\includegraphics[width=\textwidth]{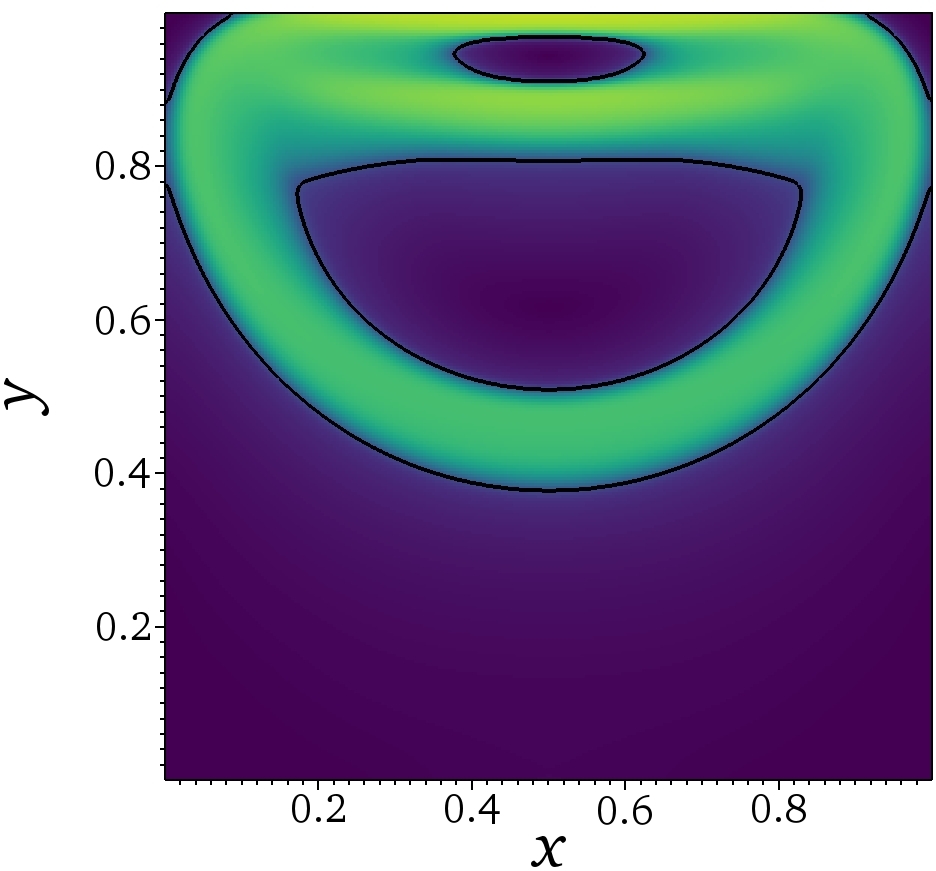}
		\caption{$\tilde{t}=6.480 \times 10^{-4}$}
	\end{subfigure}
	~
	\begin{subfigure}{0.29\textwidth}
		\includegraphics[width=\textwidth]{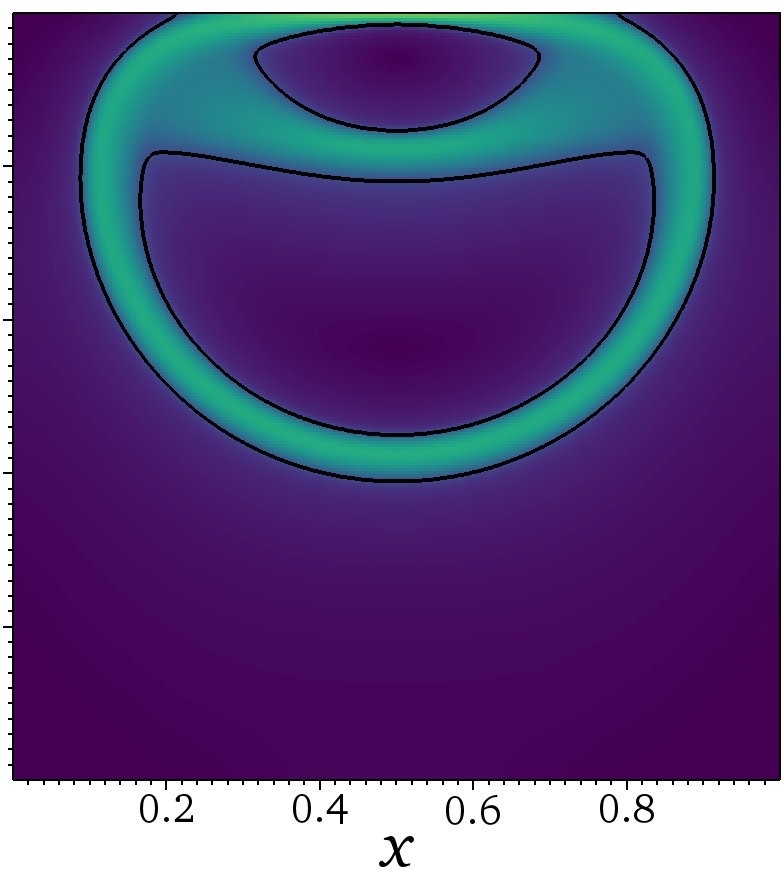}
		\caption{$\tilde{t}=1.1664 \times 10^{-3}$}
	\end{subfigure}
	~
	\begin{subfigure}{0.29\textwidth}
		\includegraphics[width=\textwidth]{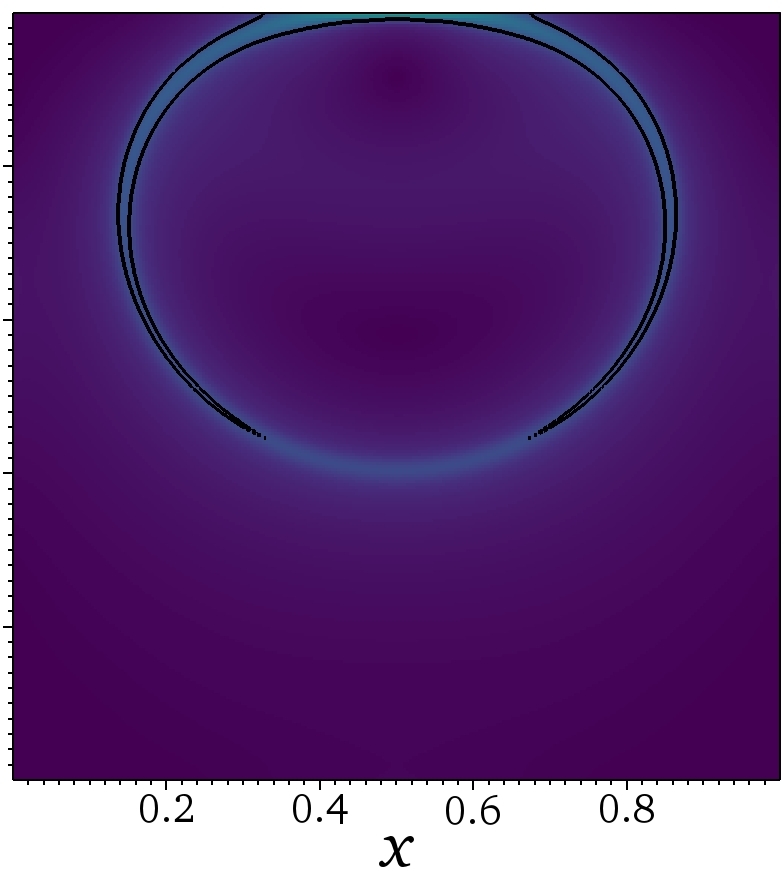}
		\caption{$\tilde{t}=1.2729 \times 10^{-3}$}
	\end{subfigure}
	\caption
	{
        Yield surface and local Reynolds number for time-dependent flow in the 2D lid-driven
        cavity. The fluid is initially at rest, and gradually transitions to a steady state
        (a)-(e).  At this point, the lid is stopped, allowing cessation in finite time (e)-(i).
	}
	\label{fig:rest-steady-cessation-2d}
\end{figure*}

Results with $\Rey=1$ and $\Bin=10$ are depicted in figure \ref{fig:rest-steady-cessation-2d}. The
yield surface is illustrated as the single black contour line $|\tensor{\tau}| =
\tau_0$,\footnote{This simple method for determining the yield surface can be improved upon for
regularised constitutive equations, see e.g.~the work of Liu, Muller and Denn
\cite{liu2003interactions}} while the heat map shows the local Reynolds number $\Rey_L$. Note that
the scaling of the heat map is logarithmic (see top of figure \ref{fig:rest-steady-cessation-2d}).

For viscoplastic fluids, unyielded regions occur in two distinct manners. Firstly, there are
so-called stagnant zones, which are connected to no-slip walls and in which the fluid has zero
velocity. On the other hand, unyielded zones with non-zero velocity are referred to as plug zones.
These are not in contact with the walls, and rotate in the interior of the cavity. In fact, these
two types of unyielded zones can never be in contact with each other, but must be separated by a
layer of yielded fluid. This is because a plug zone connected to a stagnant zone would by
association itself be stagnant, and thus could not move. 

The first five plots (a)-(e) show the transition from rest to steady-state. It is clear that the
fluid yields immediately at a very thin layer near the lid, since a large strain is applied here at
$\hat{t}=0$. Additionally, there are yielded regions which quickly propagate from the top corners
and in towards the middle, eventually combining together and leaving an unyielded plug in the upper
middle part of the cavity (a)-(c). This process occurs very quickly, in less than one percent of
the time needed to reach steady-state, but should in fact happen instantaneously, following the
discussion above. The initial connection between the stagnant zone at the bottom and the rotating
plug is a spurious artefact due to the regularisation approach. Since we regularise the apparent
viscosity, small but non-zero velocities are permissible in unyielded regions.

The steady-state shape of the resulting plug, and the unyielded region at the bottom of the cavity
(e), is characteristic for the chosen pair of Reynolds and Bingham numbers.  Comparisons to results
by Syrakos et al.~ show that the steady-state yield surfaces are indistinguishable. At this point
($\hat{t}=0.020552$), the lid is abruptly halted and held still. This removes the driving force for
the flow from the system, and leads to cessation of the flow. Since cessation occurs in finite time
for viscoplastic fluids, the cessation time $t_c$, defined as the time until the entire cavity is
in the unyielded state, is an important characteristic. In order to measure $\hat{t}_c$, we
introduce a new time $\tilde{t} = \hat{t} - 0.0205552$.  We have chosen four times during flow
cessation to compare with Syrakos et al.~\cite{syrakos2016cessation} and the flow patterns are very
similar to their published results.  Additionally, our simulations yield $\hat{t}_c = 1.2844 \times
10^{-3}$, which is a relative difference of less than $0.9\%$ compared to the previously published
results. Note also the unphysical connection between stagnant and plug zones in (i), which is
in agreement with the results and discussions of Syrakos et al.~\cite{syrakos2016cessation}. In
order to rectify these spurious artefacts, one would need a very high Papanastasiou number or use
an unregularised approach. \cite{dimakopoulos2003transient}. 


\section{Evaluation}
\label{sec:evaluation}

\subsection{Herschel-Bulkley fluids}

The results in section \ref{sec:validation} validate our code for fluids obeying power-law and
Bingham rheologies, but there are no available comparisons in the literature for Herschel-Bulkley
fluids, which exhibit power-law dependencies in addition to a non-zero yield stress. Consequently,
simulations were performed for viscoplastic fluids with the flow index $n$ equal to 0.5
and 2.0. In order to cover a wide range of flow behaviours, we also chose to vary the
Herschel-Bulkley number $\HB$ and the modified Reynolds number $\Rey^*$, as given by
\eqref{eq:reynolds_modified}. For each combination of $n$, $\HB$ and $\Rey^*$, the system was
advanced to steady-state, in order to obtain velocity profiles through the centre of the domain and
the final location of the main vortex. These results can be used as benchmark references for codes
simulating Herschel-Bulkley fluids. 

\begin{figure*}
	\includegraphics[width=\textwidth]{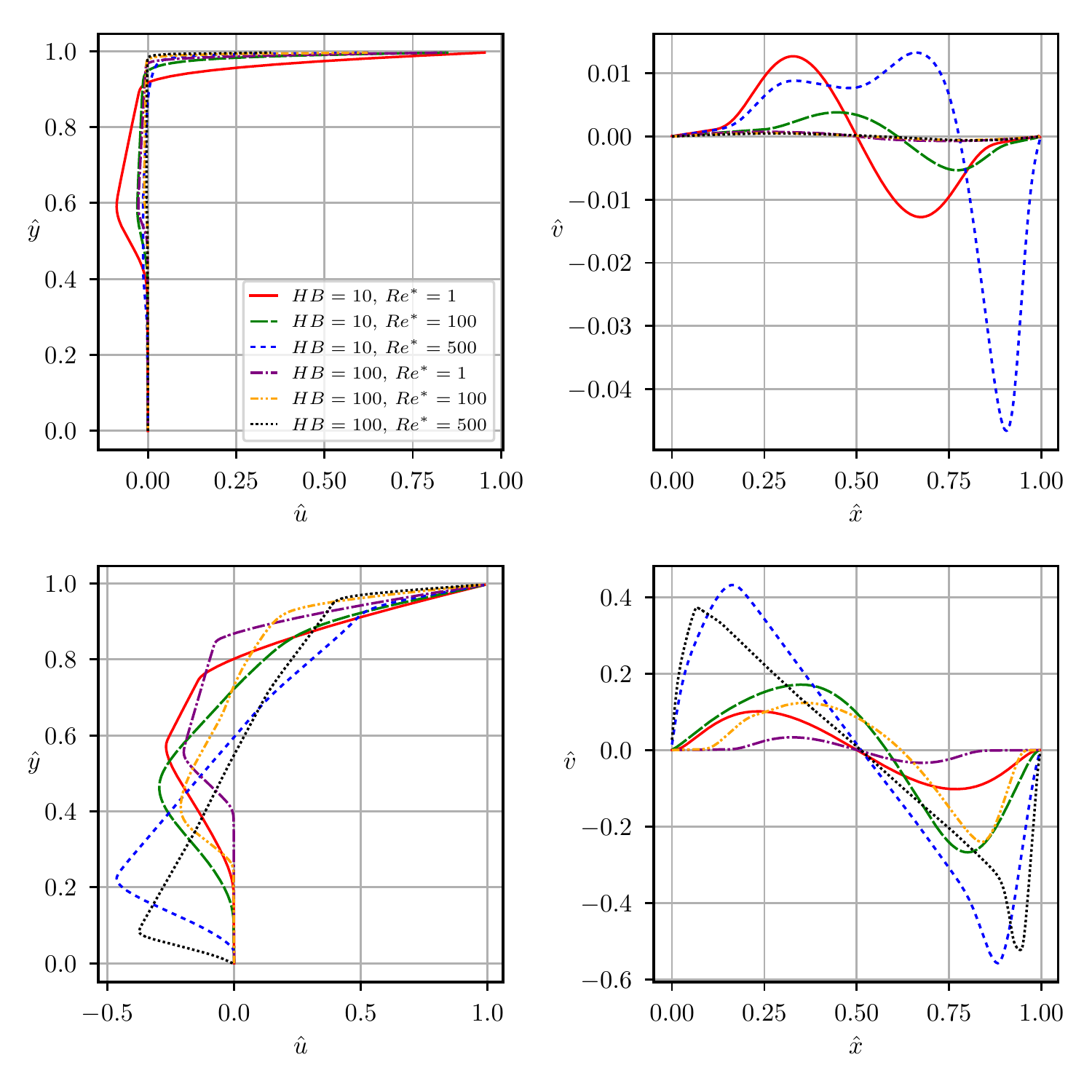}
	\caption
	{
        Steady-state velocity profiles for Herschel-Bulkley fluids with $n=0.5$ (top) and $n=2.0$
        (bottom), and various combinations of $\HB$ and $\Rey^*$, given in the legend in the top
        left plot. Note that for the dilatants with $\HB=100$, several lines almost overlap.
    }
	\label{fig:velocity_profiles_hb}
\end{figure*}

Figure \ref{fig:velocity_profiles_hb} shows the steady-state velocity profiles for each different
parameter combination. As expected, a small flow index enhances the effect of the yield stress,
resulting in weak velocity variations. This leads to the largely overlapping $x$-velocity profiles
in the upper left plot, so that the $y$-velocity profiles in the upper right one are more useful
comparisons. In the lower plots, however, where the fluid acts as a dilatant above the yield
criterion, there is a trade-off between the effects of flow index and yield stress, so that we
obtain large velocity variations throughout the middle of the domain in both directions.

\begin{figure}
	\includegraphics[width=\columnwidth]{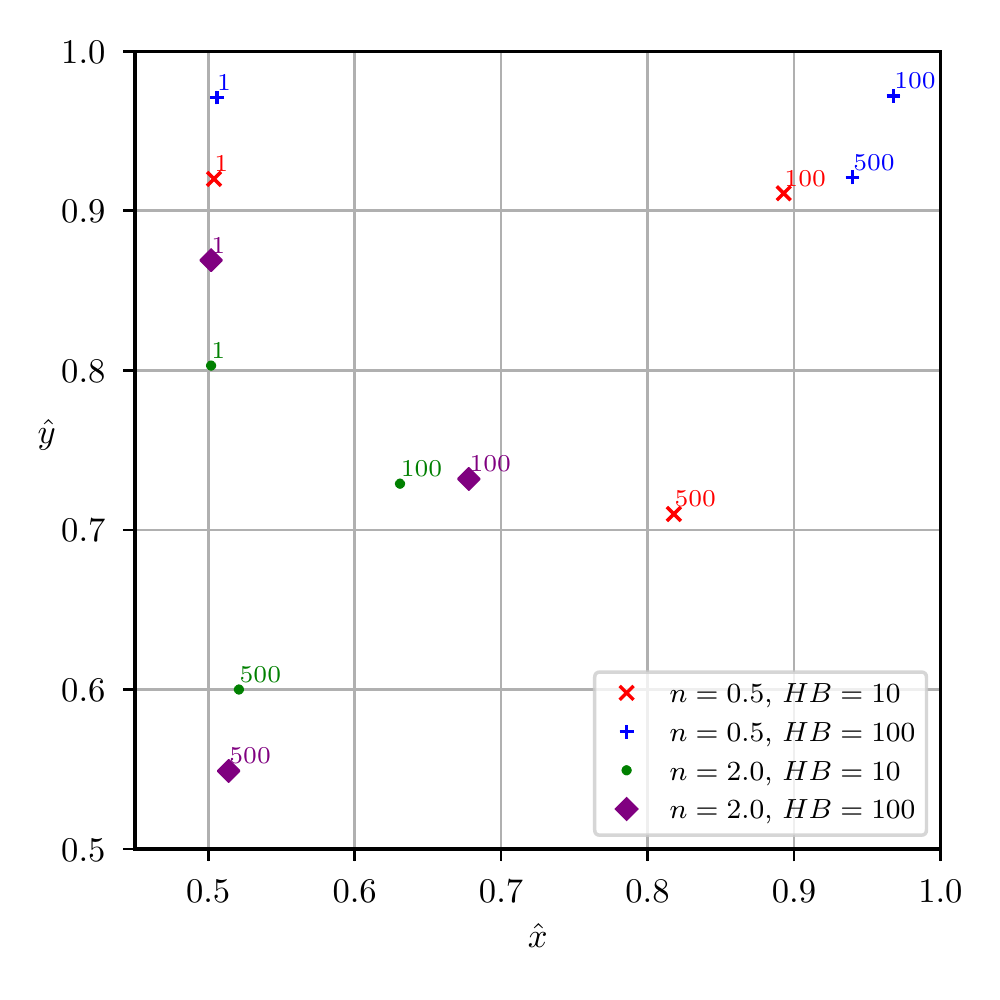}
	\caption
	{
        Steady-state vortex centres for Herschel-Bulkley fluids with different
        combinations of flow index $n$, Herschel Bulkley number $\HB$ and modified Reynolds number
        $\Rey^*$. The number next to each point is the corresponding $\Rey^*$.
    }
	\label{fig:vortex_centers_hb}
\end{figure}

In figure \ref{fig:vortex_centers_hb}, we show the steady-state locations of the main vortex
centres for each Herschel-Bulkley simulation. Comparing the results with figure
\ref{fig:vortex_centers}, we see the same effects of increasing $\HB$ as for $\Bin$, i.e.~vortex
moving upwards and to the right. Similarly, increasing $\Rey^*$ results in the vortex moving first
right and then down toward the cavity centre. Both of these observations are as expected, since the
dimensionless groups capture similar aspects of the flow. Additionally, we note that decreasing the
flow index $n$ has a similar effect as increasing the yield stress, moving the vortex centre
upwards and to the right. These additional data points serve as further benchmarking results for
Herschel-Bulkley fluids. 

\subsection{Adaptive mesh refinement}

The use of AMR improves the efficiency of our algorithm in two different ways, depending on whether
we are interested in steady-state results or transient behaviour. In the former case, we do not
care about intermediate results, and would ideally employ a large time step. Unfortunately, due to
the semi-implicit viscous solves employed in IAMR, we are unable to reach steady-state for highly
resolved simulations by iterating with large time steps. The large system requires longer runtimes
per time step, and the time step decreases proportionally with the grid size. On the other hand, we
can rapidly advance a low-resolution simulation to steady-state, and then restart the simulation
with one refinement level enabled throughout the entire domain. Due to the initial conditions, the
finer resolution will converge after a much smaller amount of iterations. Subsequently, the
simulation can be restarted again with a further level of refinement, and so on. In this manner,
the computational expense associated with time-dependent simulation to a highly resolved
steady-state is greatly reduced. As an example, advancing the 2D case with $\Rey=1$, $\Bin=10$ to
steady-state on a 16-core desktop computer took 83 minutes and 13 seconds when using a resolution
of $256^2$. By first advancing a low-resolution system with $64^2$ cells to steady-state (33
seconds), and then refining the entire domain with two levels, the runtime for obtaining the same
steady-state result is reduced to just under 15 minutes. Due to the built-in checkpoint
functionality in AMReX, restarting simulations with different runtime options in this manner is
straightforward. 

\begin{figure}
    \includegraphics[width=\columnwidth]{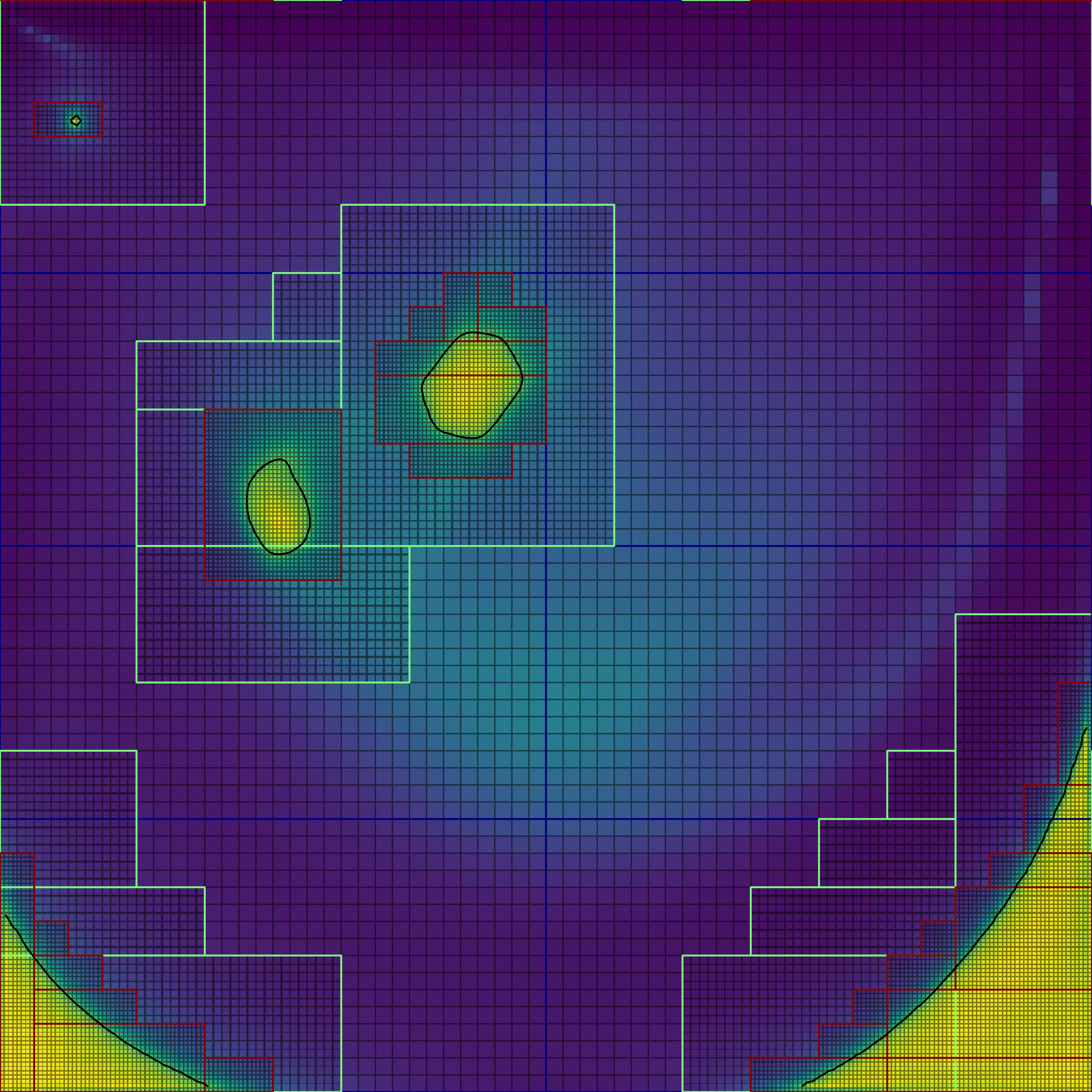}
	\caption
	{
        Illustration of spatio-temporal adaptive mesh refinement for tracking the development of
        the yield surface, with $\Rey=1000$ and $\Bin=1$. The coarsest grid discretises with cell
        spacing $1/64$, but the two refinement layers ensure that the effective resolution near the
        yield surface is $1/256$. (Multimedia view)
	}
    \label{fig:amr}
\end{figure}

In order to illustrate the usefulness of AMR in unsteady simulations, we refer to figure
\ref{fig:amr} (Multimedia view), which shows the full evolution of a Bingham fluid with $\Rey=1000$
and $\Bin=1$ from rest to steady-state with AMR enabled. The base grid consists of 64 cells in each
direction, and two levels of refinement are used to achieve high resolution near the yield surface.
This is done by tagging cells where $|\tensor{\tau}| \leq 1.05 \tau_0$ for refinement.
Consequently, we only need to use large amounts of computational resources in some regions. Due to
the block-structured AMR implementation, the grid distribution is highly parallel. Additionally,
the temporal subcycling in time ensures that we only need to decrease the step size in refined
cells -- not globally. 

\subsection{Moving to three dimensions}

\begin{figure*}
	\begin{subfigure}{0.3\textwidth}
		\includegraphics[width=\textwidth]{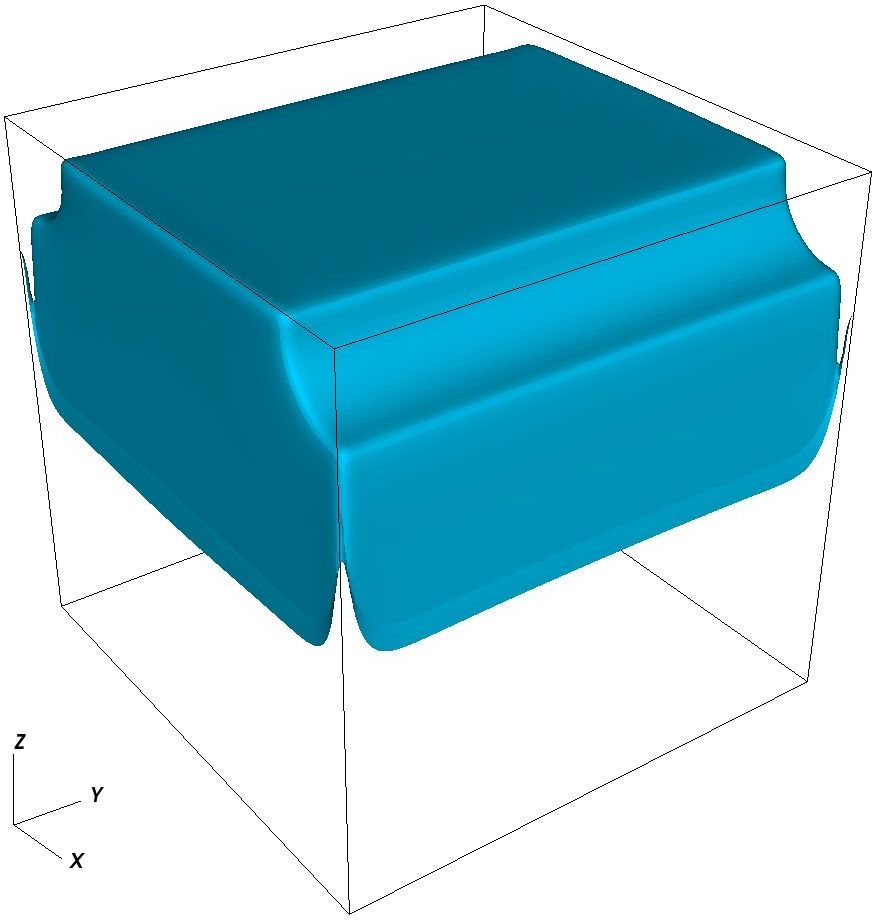}
		\caption{$\hat{t}=1.0 \times 10^{-6}$}
	\end{subfigure}
	~
	\begin{subfigure}{0.3\textwidth}
		\includegraphics[width=\textwidth]{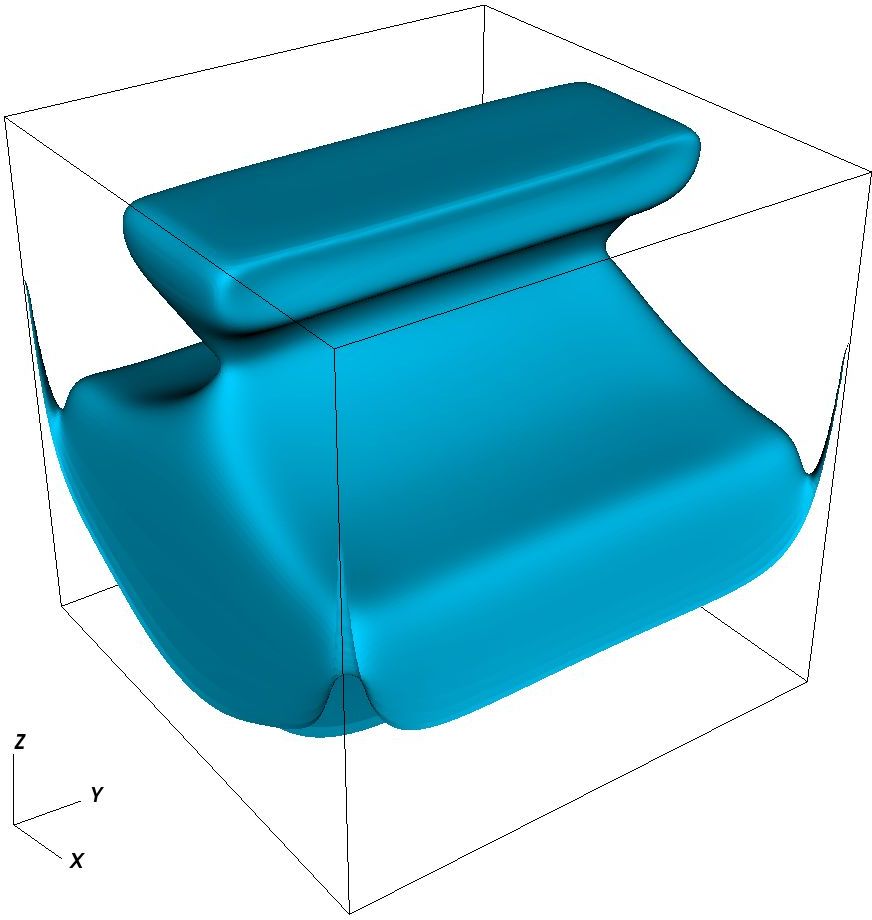}
		\caption{$\hat{t}=1.00 \times 10^{-5}$}
	\end{subfigure}
	~
	\begin{subfigure}{0.3\textwidth}
		\includegraphics[width=\textwidth]{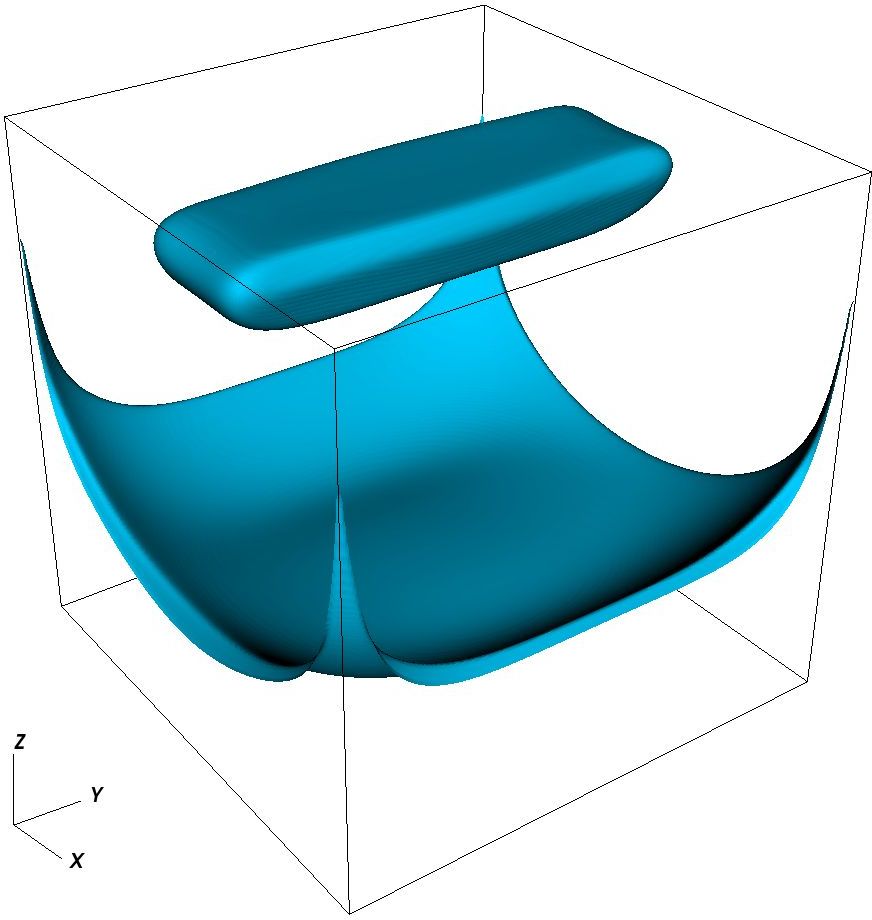}
		\caption{$\hat{t}=6.82 \times 10^{-5}$}
	\end{subfigure}
	\\[1em]
	\begin{subfigure}{0.3\textwidth}
		\includegraphics[width=\textwidth]{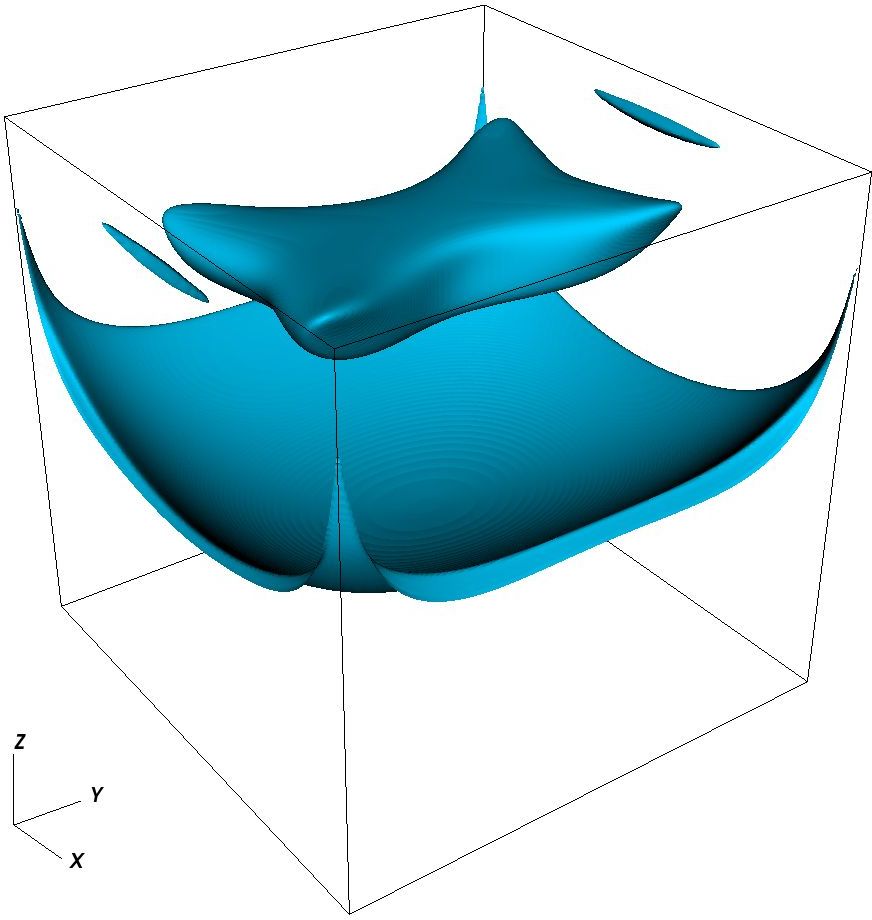}
		\caption{$\hat{t}=8.438 \times 10^{-4}$}
	\end{subfigure}
	~
	\begin{subfigure}{0.3\textwidth}
		\includegraphics[width=\textwidth]{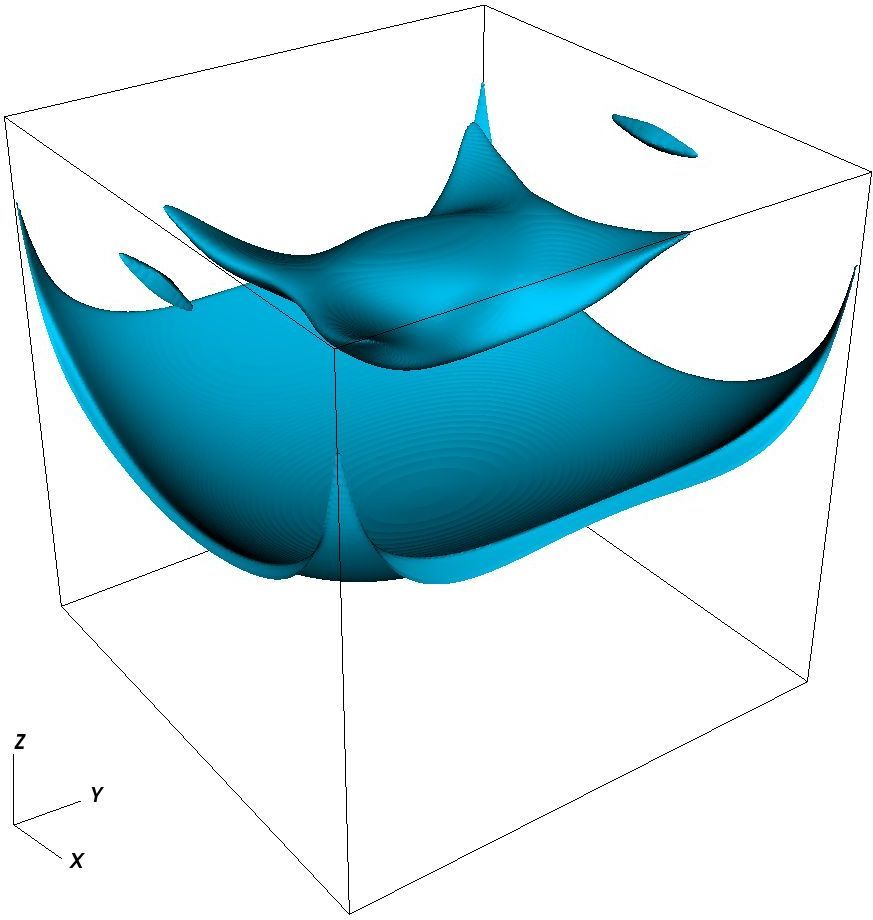}
		\caption{$\hat{t}=1.4913 \times 10^{-2}$, $\tilde{t} = 0$}
	\end{subfigure}
	~
	\begin{subfigure}{0.3\textwidth}
		\includegraphics[width=\textwidth]{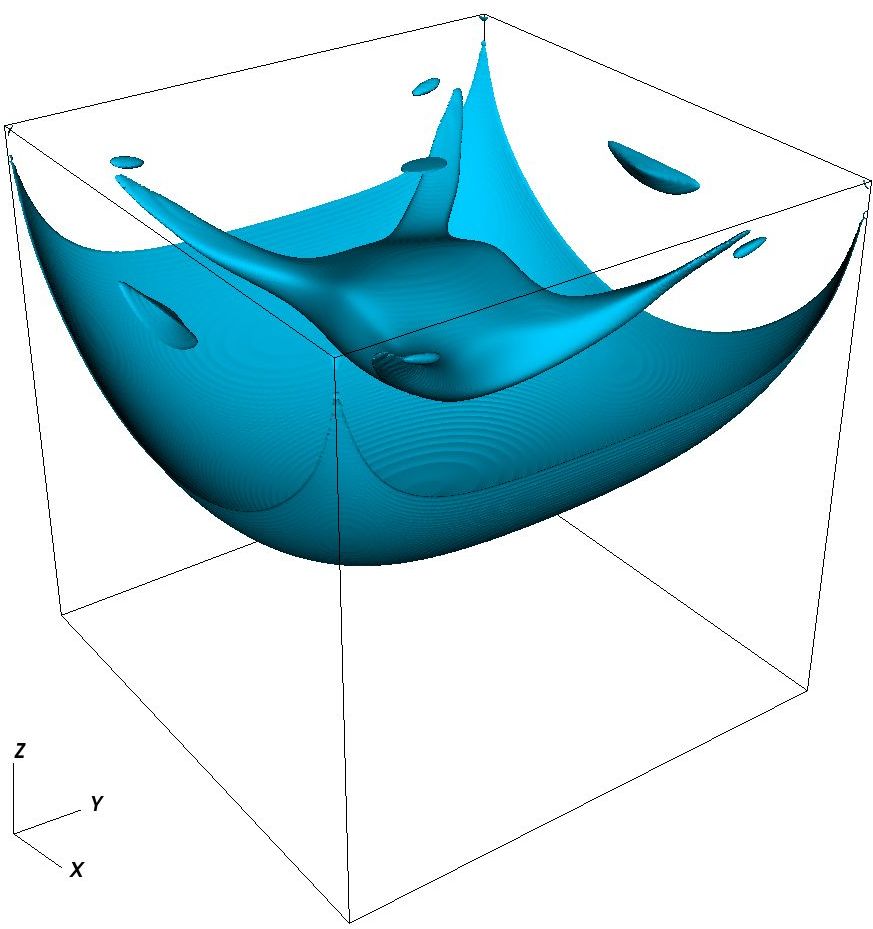}
		\caption{$\tilde{t} = 3.97 \times 10^{-5}$}
	\end{subfigure}
	\\[1em]
	\begin{subfigure}{0.3\textwidth}
		\includegraphics[width=\textwidth]{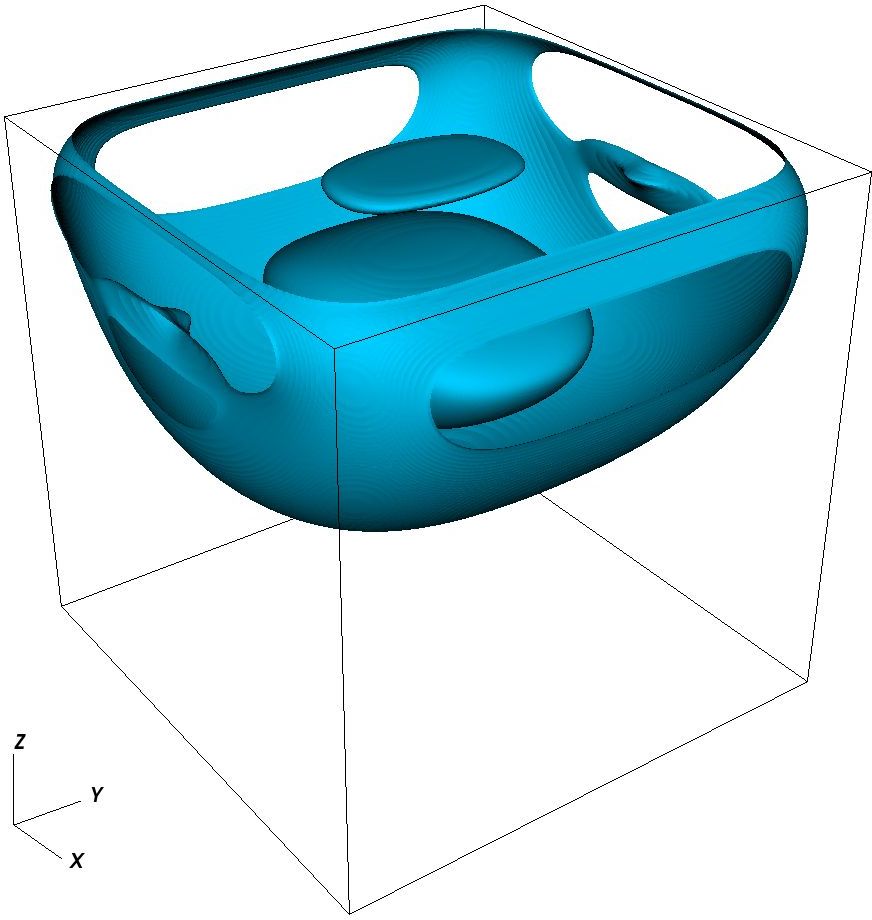}
		\caption{$\tilde{t} = 4.847 \times 10^{-4}$}
	\end{subfigure}
	~
	\begin{subfigure}{0.3\textwidth}
		\includegraphics[width=\textwidth]{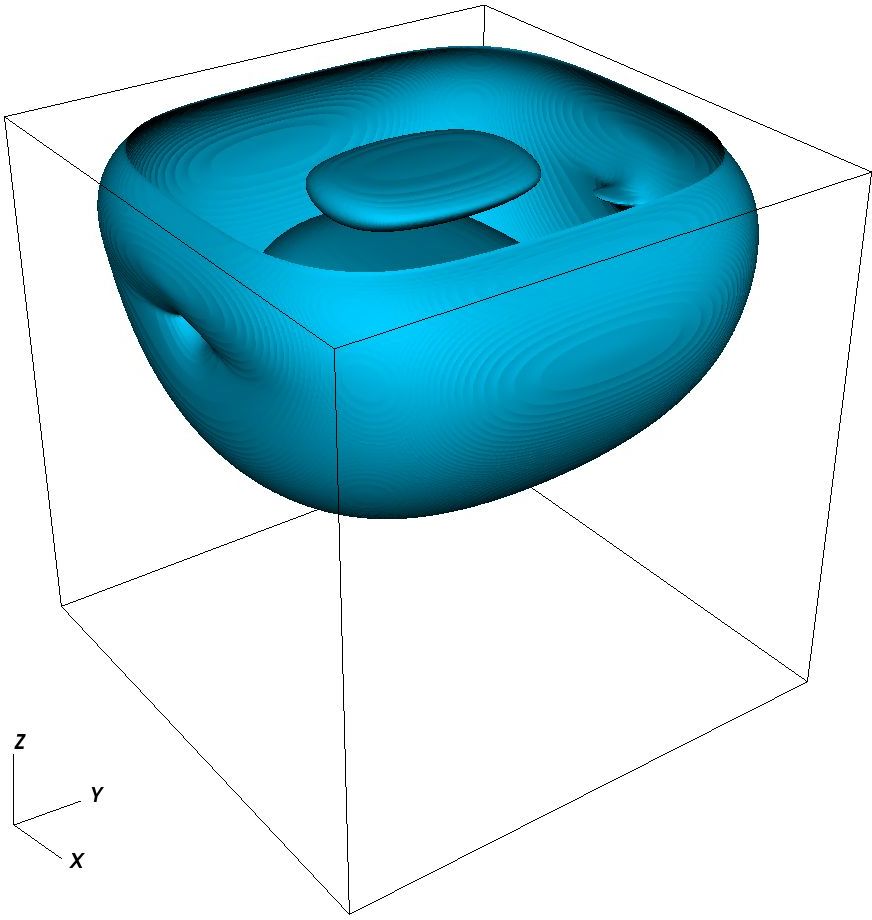}
		\caption{$\tilde{t} = 7.706 \times 10^{-4}$}
	\end{subfigure}
	~
	\begin{subfigure}{0.3\textwidth}
		\includegraphics[width=\textwidth]{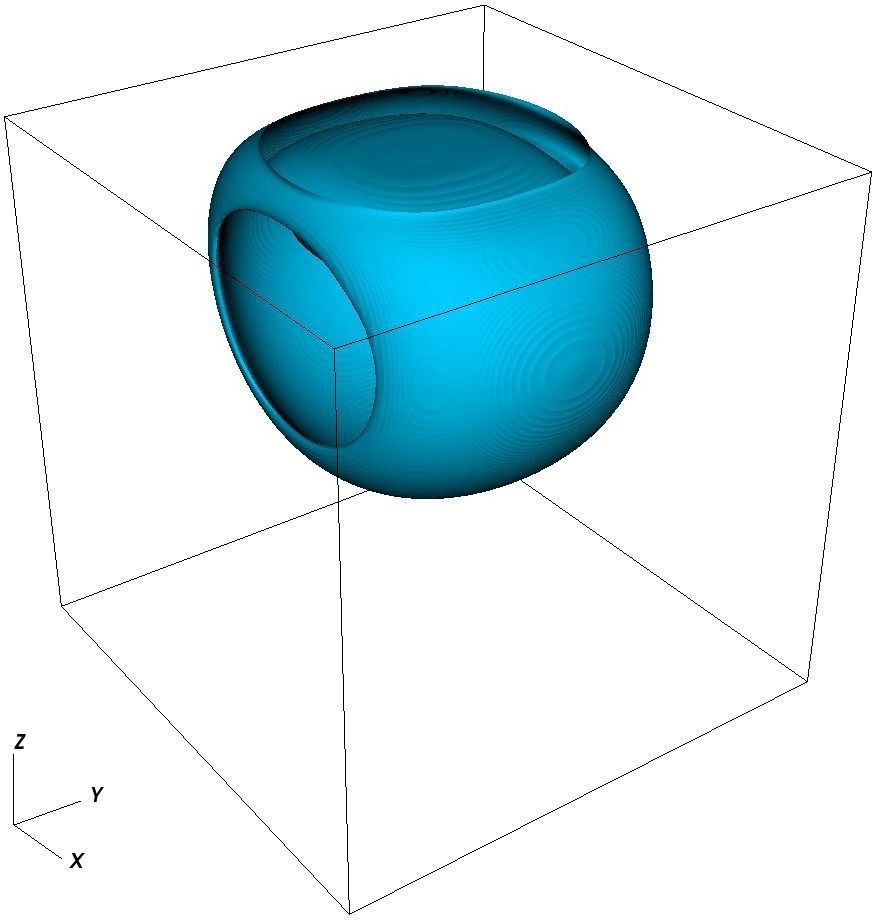}
		\caption{$\tilde{t} = 1.0564 \times 10^{-3}$}
	\end{subfigure}
	\caption
	{
        Three-dimensional time-evolution of the lid-driven cavity flow, with $\Rey = 1$ and $\Bin =
        10$. The system advances from an instantaneous start from rest to steady-state, before the
        lid motion ceases and the fluid stops. The lid moves in positive $x$-direction. The
        shapes of three-dimensional plug regions of this sort have never been identified for the
        lid-driven cavity before.
	}
	\label{fig:rest-steady-cessation-3d}
\end{figure*}

Since the augmented IAMR code is efficient and parallelises well, it is possible to run unsteady
simulations of fully three-dimensional viscoplastic flows. As an extension of the lid-driven cavity
for Bingham fluids in 2D, we consider a unit cube with stationary no-slip conditions on the floor
and all four walls, while the lid (at $\hat{z} = 1$) drives the flow. 
We expect the results of such a simulation to be similar to the classical
two-dimensional case, which means that we can compare the two and expect similar flow
characteristics. However, we would only get identical results by using periodic boundary condition
on the walls which are parallel to the flow vortices. Since we apply no-slip conditions, there will
be three-dimensional effects on the results, meaning that we can use them as separate reference
results.
The spatial domain is discretised using 256 cells in each directions, i.e.~16,777,216 cells in
total. As before, the system is initially at rest. When steady-state is reached, we stop the motion
of the lid and allow cessation to take place. Figure \ref{fig:rest-steady-cessation-3d} illustrates
how the yield surface evolves through time.  Note that it is still computed as the contour
$|\tensor{\tau}| = \tau_0$, but in 3D we can actually visualise it as a surface. The snapshot times
in figure \ref{fig:rest-steady-cessation-3d} are similar to those used in figure
\ref{fig:rest-steady-cessation-2d}, but vary by small factors. Note also here the presence of
numerical artefacts due to regularisation, demonstrated by connected stagnant and plug zones.
Qualitatively, the resemblance with the 2D case is striking, as the flow clearly goes through the
same steps to reach steady-state and cessation. Having said that, the third dimension allows a
richer picture of the shape of the yield surface. Especially fascinating is the importance of the
four vertical edges, which play a similar role as the 2D corners, and lead to the yield surface
stretching out in four directions from the centre. 
It would not be possible to capture this type of yield surface shape without
performing fully 3D fluid simulations.
We emphasise that this is the first time fully three-dimensional yield surfaces have been presented
for the lid-driven cavity case.  Previously, the only 3D results are slices of rigid zones
presented by Olshanskii \cite{olshanskii2009analysis} (finite differences, 262,144 cells), and the
heatmap slices presented in the bi-viscosity simulations performed by Elias et
al.~\cite{elias2006parallel}, (819,468 elements).

\begin{figure*}
	\begin{subfigure}{0.3\textwidth}
		\includegraphics[width=\textwidth]{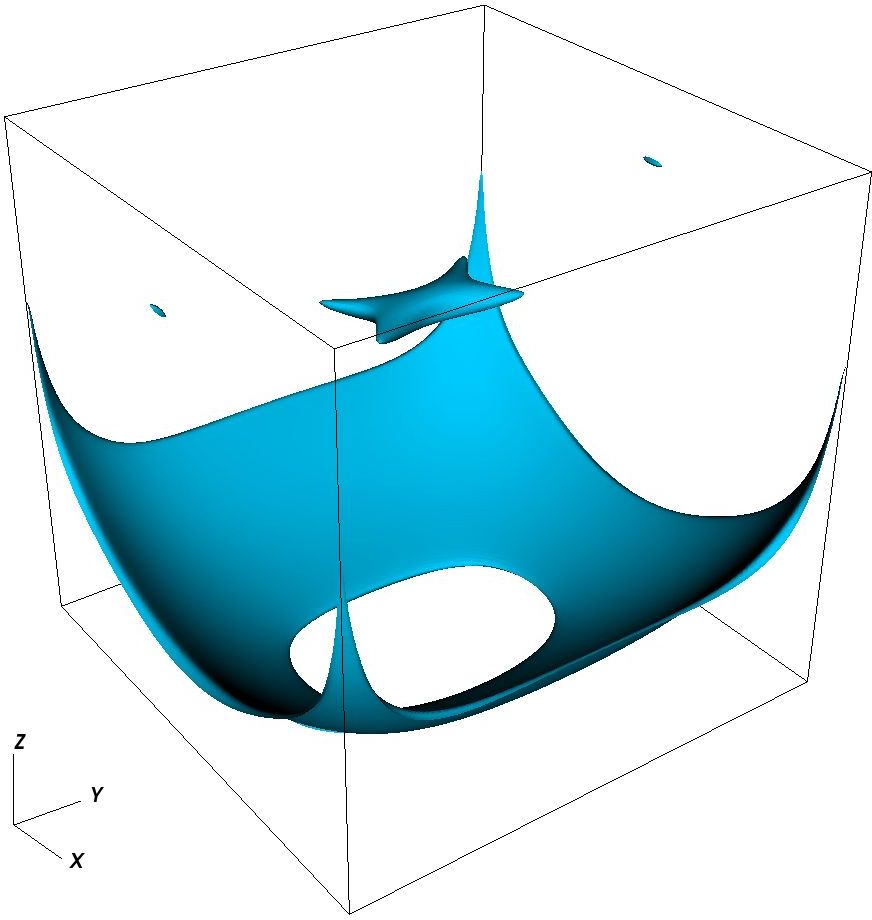}
		\caption{$Re=1,\Bin=1$}
	\end{subfigure}
	\hspace{1em}
	\begin{subfigure}{0.3\textwidth}
		\includegraphics[width=\textwidth]{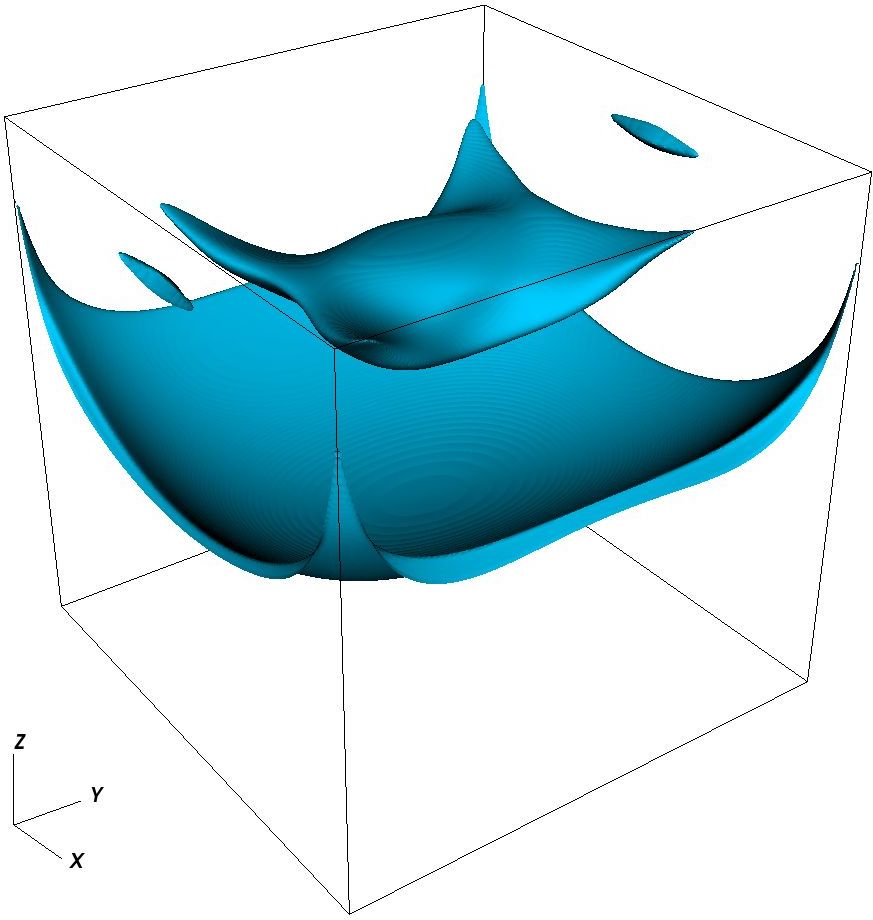}
		\caption{$Re=1, \Bin=10$}
	\end{subfigure}
	\hspace{1em}
	\begin{subfigure}{0.3\textwidth}
		\includegraphics[width=\textwidth]{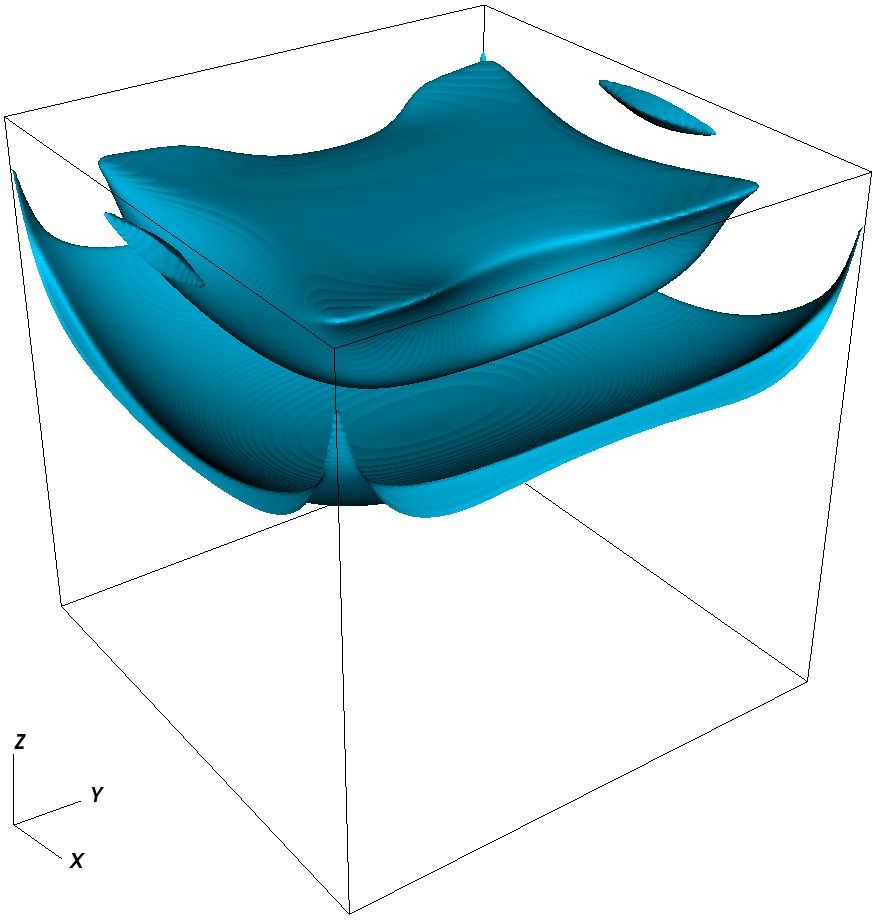}
		\caption{$Re=1, \Bin=100$}
	\end{subfigure}
	~
	\begin{subfigure}{0.3\textwidth}
		\includegraphics[width=\textwidth]{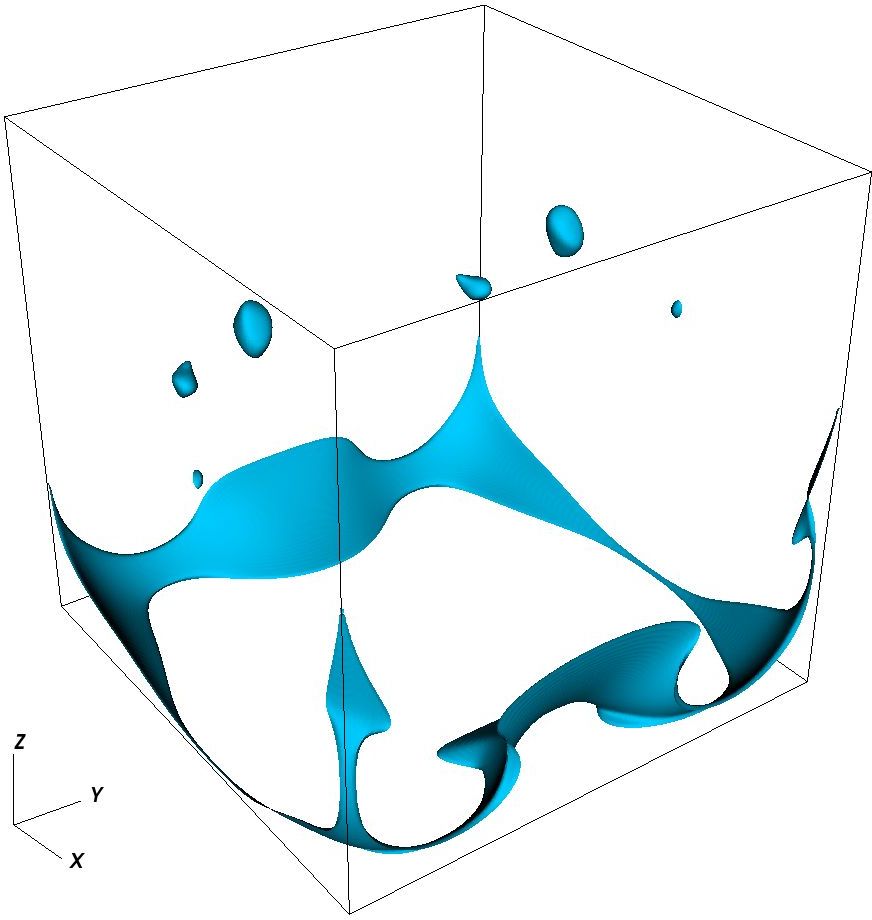}
		\caption{$Re=1000, \Bin=1$}
	\end{subfigure}
	\hspace{1em}
	\begin{subfigure}{0.3\textwidth}
		\includegraphics[width=\textwidth]{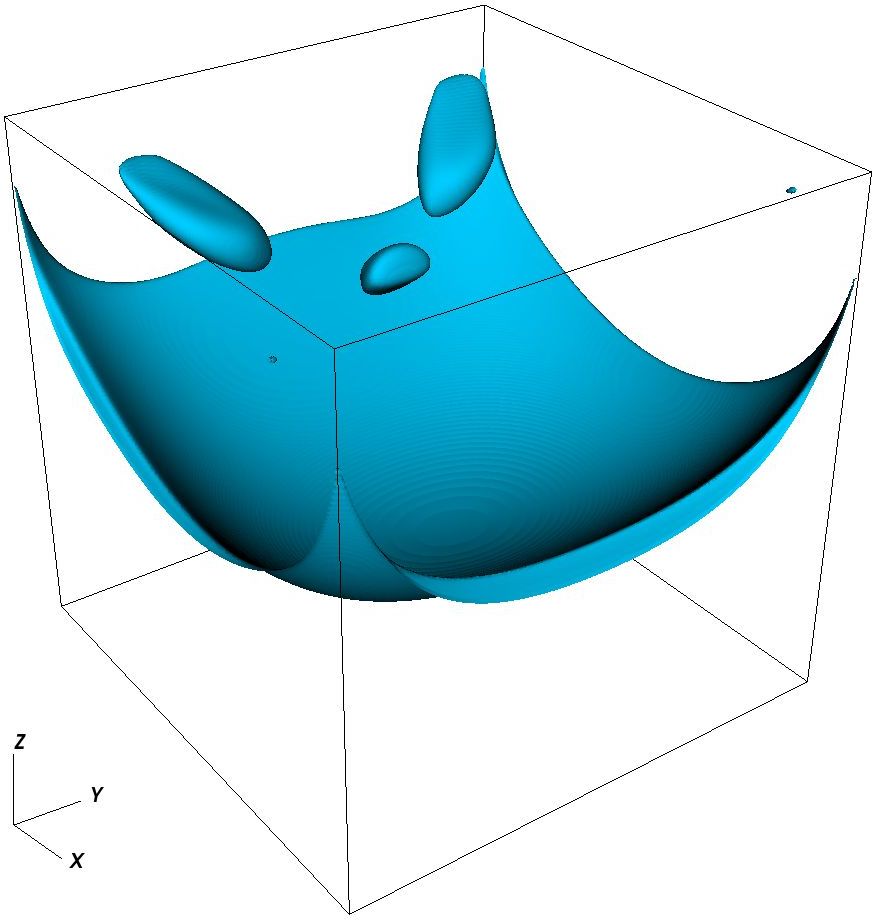}
		\caption{$Re=1000, \Bin=10$}
	\end{subfigure}
	\hspace{1em}
	\begin{subfigure}{0.3\textwidth}
		\includegraphics[width=\textwidth]{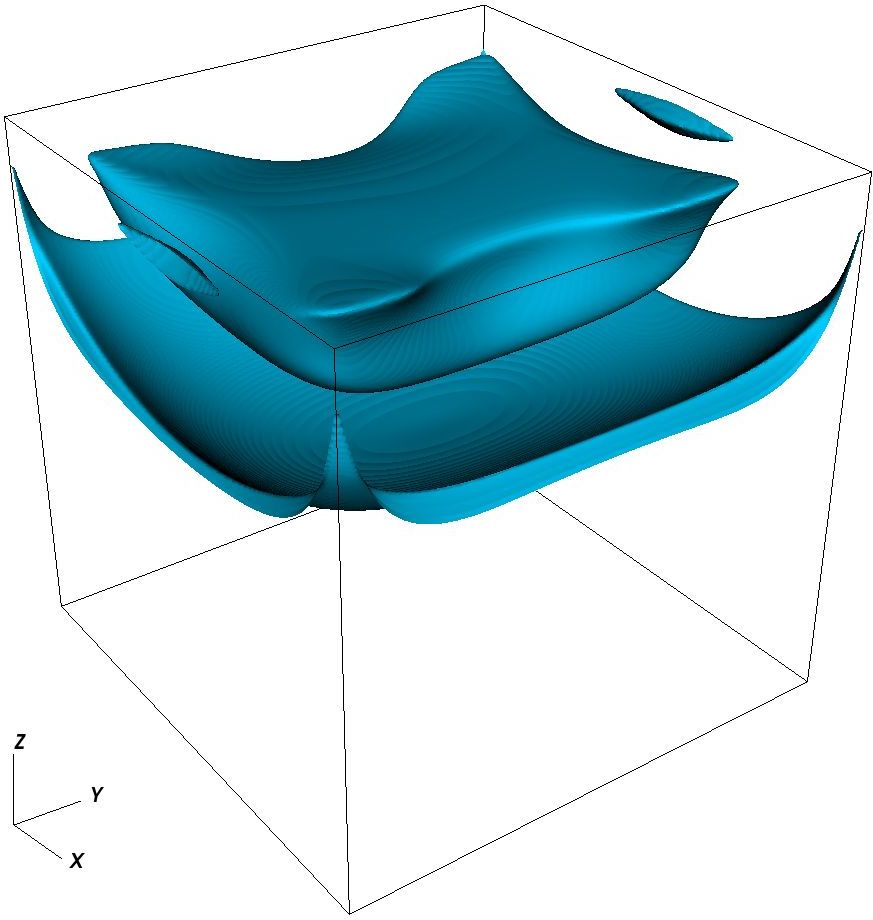}
		\caption{$Re=1000, \Bin=100$}
	\end{subfigure}
	\caption
	{
		Steady-state yield surface for Bingham fluids with various Reynolds and Bingham numbers in
		the three-dimensional lid-driven cavity. The lid moves in positive $x$-direction.
	}
	\label{fig:steady-3d}
\end{figure*}

In figure \ref{fig:steady-3d}, steady-state yield surfaces are shown for various combinations of
$\Rey$ and $\Bin$. Readers who are familiar with the two-dimensional test case (see e.g.~Syrakos et
al.\cite{syrakos2014performance}) will recognise the similarity between those results and slices
through these three-dimensional regions. Note that the slices are taken at a slight angle in the
$xy$-plane. These simulations illustrate the code's ability to solve the governing PDEs for
time-dependent three-dimensional flows of non-Newtonian fluids following the Herschel-Bulkley
model.

\begin{figure*}
  \includegraphics[width=\textwidth]{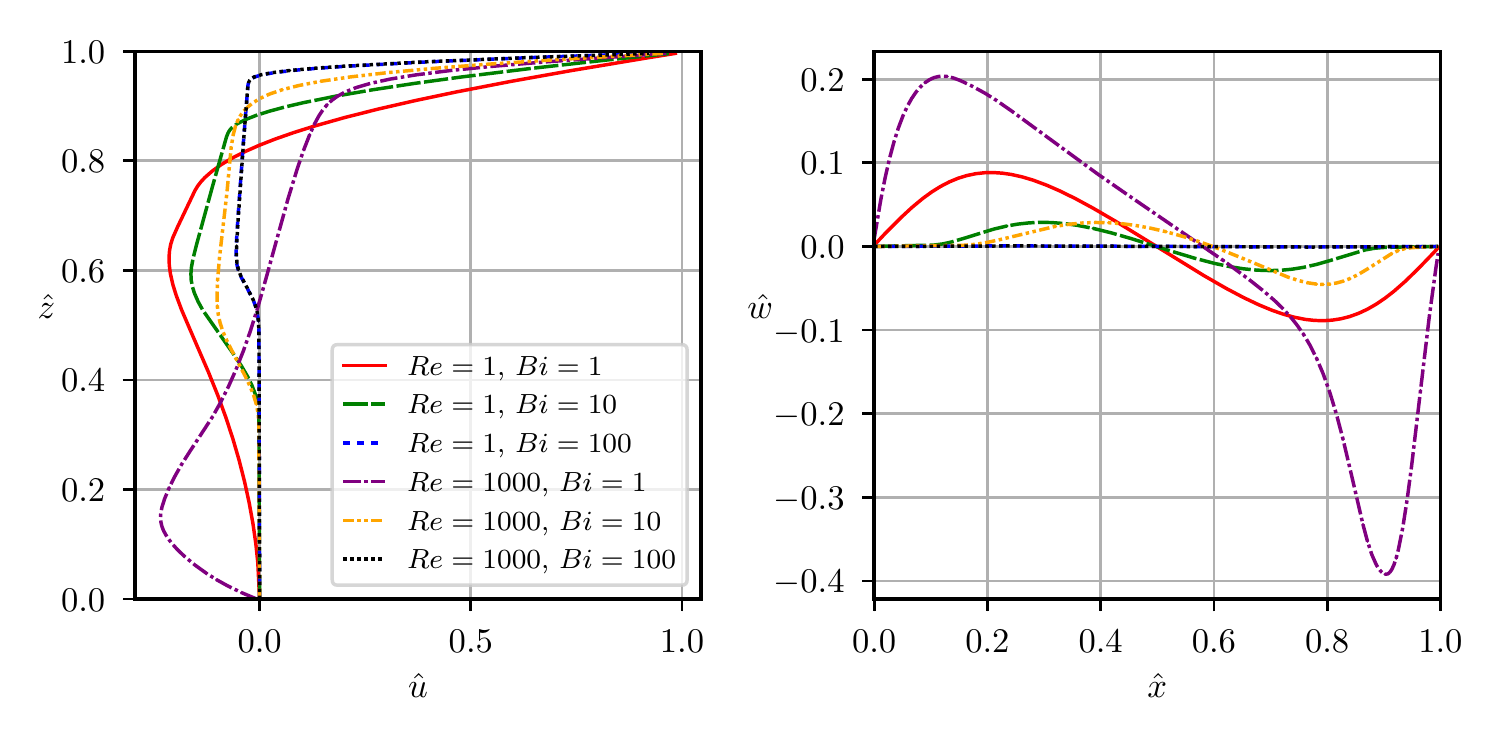}
  \caption
  {
      Velocity profiles in the centre of the three-dimensional cavity, provided as reference
      solutions for 3D viscoplastic solvers. The effect of Reynolds and Bingham numbers is clear.
      Note that the two cases with $Bi=100$ are indistinguishable.
  }
  \label{fig:velocity_profiles_3d}
\end{figure*}

While the shapes of the three-dimensional yield surfaces shown in figures
\ref{fig:rest-steady-cessation-3d} and \ref{fig:steady-3d} provide interesting insight about the
three-dimensional effects, they do not provide quantitative comparisons for other codes for
three-dimensional viscoplastic fluid flows. Consequently, we show the corresponding steady-state
velocity profiles in figure \ref{fig:velocity_profiles_3d}. The first velocity component is
considered on the line $\hat{x}=0.5$, $\hat{y}=0.5$, while the $x$- and $z$-directions are swapped
for the third component. As is evident, there are notable similarities between the two-dimensional
simulations, as expected. These profiles can act as benchmark results for future yield stress
simulation codes.

\subsection{Parallel performance}

Among the most impressive attributes of codes built within the AMReX framework, is the intrinsic
scalability on state-of-the-art computer architectures.  Due to major advances within parallel
computing in the last few decades, accurate measurement of how well-suited algorithms are for
parallel processing has received a lot of attention. Moreland and Oldfield provide an excellent
overview of contemporary formal metrics for such measurements \cite{moreland2015formal}.

Suppose we are trying to solve a computational problem of size $N$. Then the speed-up of the
parallel algorithm is given by the ratio of the parallel execution time to that of the best serial
algorithm. The theoretically best possible speed-up occurs if the entire task can be divided into
as many chunks of equal work as there are cores being used, so that it is distributed among the
available cores without any overhead. In this utopian case, we have speed-up equal to the processor
core count.  Although speed-up is a useful metric for analysing the parallel performance of an
algorithm, it has one caveat. Since the serial runtime is necessary in order to compute them, it
quickly becomes impractical for analysis of large-scale applications. Computing the serial runtime
for very large $N$ takes far too long. Scaling where the problem size is kept fixed, and the amount
of allocated resources is increased, is known as strong scaling. On the other hand, the problem
size can be scaled so that it is proportional to the core amount. If one core is used to solve the
problem with size $N^*$, one would then ideally expect the parallel runtime of a problem with size
$PN^*$ to remain constant when deployed on $P$ cores.  This type of analysis is called weak
scaling. Finally, it may in some case be difficult to keep the problem size proportional to the job
size, so that weak scaling analysis becomes difficult to obtain. However, one can measure the
amount of work done per time, referred to as the rate. The rate is computed as the ratio of problem
size to parallel runtime, and thus removes the weak scaling dependency between problem size and
core amount. Sampling the rate across a selection of problem sizes and core amounts can thus help
the simulation runner pick runtime parameters which yield a high computational efficiency.

Our main interest lies in ensuring that the scalability is unaffected by the extension to
generalised Newtonian fluids. As such, we run a test-suite with varying problem sizes and processor
cores for the 3D lid-driven cavity test when the domain is filled by a Bingham fluid with $\Rey =
1$ and $\Bin = 10$. In order to avoid the problems associated with strong and weak scaling, we vary
the number of cores and problem resolution, but only use few cores for smaller problems and many
cores for larger ones.  Subsequently, we compute the rate for the cell updates, which gives proper
insight into the resulting computational efficiency. In order to evaluate the parallel performance
rather than the effect of system state, we use a constant time step, advect the system by 100 time
steps, and measure the average runtime per time step for the next ten.

\begin{figure*}
  \includegraphics[width=\textwidth]{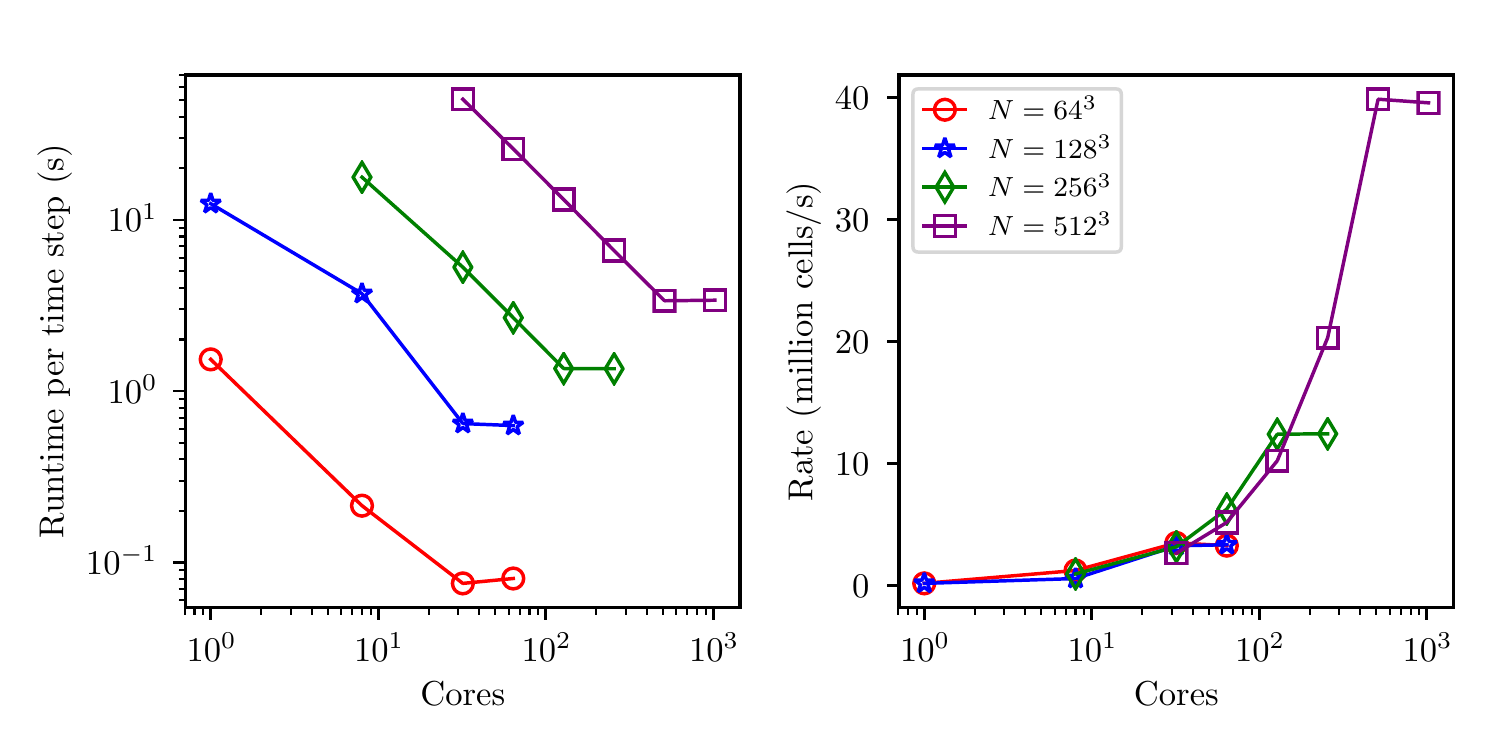}
  \caption
  {
    Strong scaling results for a Bingham-fluid in the lid-driven cavity. The left plot shows the
    reduction in runtime per time step as the amount of processor cores is increased, for various
    problem sizes. The right plot shows how the corresponding computation rate grows. Note that
    both axes are logarithmic in the left plot, while only the primary axis employs logarithmic
    scaling in the right one.
  }
  \label{fig:csd3_strong_scaling}
\end{figure*}

Figure \ref{fig:csd3_strong_scaling} illustrates the results of average runtimes and rate for the
scaling test. For a given problem size, the reduction in runtime (and increase in rate) is
basically optimal until a core limit is reached. At this point, the MPI communication overhead
becomes too large for further parallelisation, since the grid has already been broken down too far.
However, as the problem gets bigger this core limit naturally increases, so that the computational
rate can continue increasing as long as larger problems are solved. When simulating at a given
problem resolution, it is therefore best to choose the lowest amount of cores which maximise the
rate. Note that these tests do not utilise OpenMP tiling and have not been optimised in terms of
runtime parameters for parallelisation (such as maximum size for subgrids), and still the
scalability is excellent.

\section{Conclusions}
\label{sec:conclusions}

The publicly available software suite IAMR facilitates rapid simulations of incompressible fluids
governed by the Navier-Stokes equations in a structured, adaptive framework with state-of-the-art
parallelisation on contemporary supercomputer architectures. By augmenting this code, originally
designed for Newtonian flows, we allow it to deal with generalised Newtonian fluids using a
regularised Herschel-Bulkley model. In doing so, we provide a code capable of simulating
large-scale viscoplastic fluid systems efficiently enough to investigate three-dimensional,
time-dependent systems. Thorough validation is performed through comparisons with results from
peer-reviewed studies of the lid-driven cavity problem, and these benchmark results are expanded
upon through a wider range of Reynolds and Bingham numbers. Finally, we have evaluated the ability
of the code to deal with Herschel-Bulkley fluids, and investigated three-dimensional yield surfaces
in the lid-driven cavity. For both Herschel-Bulkley fluids and Bingham fluids in three dimensions,
reference results are provided for the first time.



\begin{acknowledgments}
K.~S.~would like to acknowledge the EPSRC Centre for Doctoral Training in Computational
Methods for Materials Science for funding under grant number EP/L015552/1. Additionally, he
acknowledges the funding and technical support from BP through the BP International Centre for
Advanced Materials (BP-ICAM) which made this research possible.
\end{acknowledgments}

\bibliography{../References}

\end{document}